\documentclass{article}
\usepackage{amssymb}

\usepackage{graphicx}
\usepackage{amsmath}

\input{tcilatex}

\begin{document}

\vskip 0.5cm
\begin{center}
{\bf COMPLEX ANGULAR MOMENTUM DIAGONALIZATION}
\centerline{\bf OF THE BETHE-SALPETER STRUCTURE} 
\centerline{\bf IN GENERAL QUANTUM FIELD THEORY}

\vskip 0.3cm
\centerline{\bf J. Bros} 

\vskip 0.2cm
\centerline{Service de Physique Th\'eorique, CEA-Saclay}
\centerline{F-91191, Gif-sur-Yvette Cedex, FRANCE}

\vskip 0.3cm
\centerline{\bf G.A. Viano}  

\vskip 0.2cm
\centerline{Istituto Nazionale di Fisica Nucleare, Sezione di Genova,} 
\centerline{Dipartimento di Fisica dell'Universit\`a di Genova,} 
\centerline{I-16146 Genova, ITALY}

\vskip 0.5cm

\vskip 0.3cm%

\textbf{Abstract}
\end{center}

The Complex Angular Momentum (CAM) representation of (scalar) four-point
functions has been previously established starting from the general
principles of local relativistic Quantum Field Theory (QFT). Here, we carry
out the diagonalization of the general $t$-channel Bethe--Salpeter (BS) structure
of four point functions in the corresponding CAM variable $\lambda _{t},$
for all negative values of the squared-energy variable $t$.  
This diagonalization is closely related to the existence of
BS-equations for the absorptive parts in the crossed channels, interpreted as 
convolution equations with spectral properties.    
The production of Regge poles equipped with factorized residues involving
Euclidean three-point functions appears as conceptually built-in in the
analytic
axiomatic framework of QFT. The existence of leading Reggeon terms 
governing
the asymptotic behaviour of the four-point function at fixed $t$ is
strictly
conditioned by the asymptotic behaviour of a global Bethe-Salpeter
kernel
of the theory.

\section{Introduction}

In a previous paper [1], we have proved that the existence 
of analytic structures in the 
Complex Angular Momentum (CAM) variables is a general property of 
Quantum Field Theories (QFT) satisfying the basic principles [2] of 
local commutativity, Poincar\'e invariance, spectral condition and 
temperate ultraviolet behaviour. 
Considering four-point functions 
$H([k])$ 
of general relativistic scalar fields in a given two-field
channel called the ``$t-$channel'', we have in fact shown that appropriate 
Laplace-type transforms $\tilde H^{(s)}$ and 
$\tilde H^{(a)}$ of $H$ are 
{\sl holomorphic functions of a CAM variable} $\lambda_t$, 
dual to the (off-shell) scattering angle 
$\Theta_t$ of the $t-$channel. The analyticity 
domain which was obtained for  
$\tilde H^{(s)}$ and 
$\tilde H^{(a)}$ 
is a half-plane of the form 
$\Re e \lambda_t > N_H$, 
where the number $N_H$ ($N_H >0$) corresponds to a certain ``degree of
temperateness'' of  
$H([k])$ at large momenta. This domain was obtained for all negative values of the squared
total energy $t$ of the channel considered.

\vskip 0.2cm  
It is the purpose of our program to 
investigate under which general conditions the
transforms $\tilde H^{(s)}$ and  $\tilde H^{(a)}$ of $H$ 
may admit meromorpic
continuations in a domain of the joint complex variables $(t,
\lambda_t)$
which constitutes a ``bridge'' between  
a) the ``primitive set ''  $\{(t,\lambda_t); \  t<0;\   {\rm Re }
\ \lambda_t >N_H\}$
obtained in [1] (and from which poles are excluded), and 
b) a complex neighbourhood of a real set of the form
$\{(t,\lambda_t);\  0<t<t_0;\  \lambda_t  > N(t)\}$
in which possible poles of $\tilde H^{(s)}$ and $\tilde H^{(a)}$ in the
variables $(t,\lambda_t)$ (namely ``Regge poles'') might interpolate
bound states of $H$, supposed to be present in that real set (at 
integral values $\lambda_t = 0, 1,\ldots$).

In the framework of analyticity properties implied by the general
principles of
QFT,  the existence of a meromorphic continuation of $H$ exhibiting real
poles
at $t>0$ under the two-particle threshold (resp. complex poles in the
two-particle
second sheet of the $t-$plane) is a consequence of the general
Bethe-Salpeter
structure of QFT, in which the additional postulate of  ``Asymptotic
Completeness''
(or ``off-shell unitarity'') plays an essential role [8,9].

As we have announced it in [3], we shall
show that the general Bethe-Salpeter (BS) structure
of QFT is
also operational for displaying the {\sl Regge pole interpolation of
bound states}
as a consequence of the basic principles of QFT.  In order to perform
this program,
it is necessary

i) to extend the analysis of the general BS structure of field theory so
as to include
the complex angular momentum variables, 
which requires in a first step that we restrict
ourselves to the
range $t<0$, along the line of [1];

ii) to justify the analytic continuation of this structure at $t>0$.

\vskip 0.2cm

Part i) is the object of the present paper, while part ii) will be
treated in a further
paper.

The contents of this paper can be described as follows.  
We start from a $t-$channel Bethe--Salpeter type 
structural equation for the four-point
function of scalar fields; the latter is 
considered in the space of Euclidean   
energy-momenta
which is fully contained in the primitive axiomatic analyticity domain. 
Such an integral equation involves a 
(``regularized two-particle-irreducible'') 
Bethe-Salpeter-type kernel $B$ 
whose introduction is completely justified 
on the basis of the general principles 
of local QFT and (as proved in [8,9])  
the complete two-field structure of the theory in the $t-$channel 
is encoded in this equation.  
These results and the 
corresponding structural equation for the absorptive parts 
in the crossed $s$ and $u-$channels [8,10]
are recalled in Sec 2. 
In Sec.3, we work out various aspects of these 
structures in the mass variables and angular 
variables and ultimately in the CAM variable $\lambda_t$. 
After having written the BS-equation in the mass variables and angular
variables we perform its partial-wave decomposition in the Euclidean
region in terms of ``off-shell Euclidean partial-waves''. 
Next we are interested in performing the 
analytic continuation of the BS-equation 
from Euclidean subspace to
the real Minkowskian subspace {\sl by travelling 
in the complex angular variables
of the $t-$channel inside the ``enlarged analyticity domain'' 
obtained in} [1]  
until reaching the spectral sets 
of the crossed $s$ and $u-$channels.  
This allows one to give a new approach to the ($t-$channel) BS-equations  
for the symmetric and antisymmetric combinations of the
$s$ and $u-$channel absorptive parts.
In this study, an important role is played by 
an appropriate class of holomorphic kernels 
on the complexified hyperboloid, 
called ``perikernels'', and by the ``$\star^{(c)}-$convolution 
product'' of such kernels [11], which is diagonalized by the 
Laplace-type transformation $L_d$ introduced in [1]. 
It follows 
that the BS-equations for the  
$s$ and $u-$channel absorptive parts
admit ``$t$-channel Laplace-type
transforms'' which are themseves Fredholm-type integral equations 
depending 
analytically of the CAM  
variable $\lambda _{t}$ in a half-plane of the form $\Re e \lambda _{t}>N$. 
In view of the results of [1] 
(``Froissart-Gribov equalities''), these integral relations  
provide Carlsonian
interpolations {\sl in this half-plane} of the corresponding  
BS-equations for the even and odd Euclidean partial waves. 
In the subsequent part of this paper  (Sec 4), the theory of Fredholm
resolvent
equations and specifically the 
${\cal N} \over {\cal D}$ method, ${\cal N}$ and  
${\cal D}$ being regarded as analytic functions of both variables $t$
and $\lambda_t$,
are shown to provide a general framework for generating Regge poles
equipped
with factorized residues involving Euclidean three-point functions.
The existence of corresponding leading ``Reggeon terms'' in the
asymptotic
behaviour of $H$ at large $s$ and fixed values of $t$ 
and of the mass variables necessitates the meromorphic
continuation of
$\tilde H^{(s)}$ and $\tilde H^{(a)}$ in a strip of the form
$N_B(t) < {\rm Re} \ \lambda_t < N_H$: this property is 
shown to be strictly conditioned
by the existence of a ``two-particle irreducible kernel'' $B$ whose
behaviour
in the complex $s$ (or $u$) plane at fixed values of $t$ and of the mass
variables
is bounded by $s^{N_B(t)}$, with $N_B(t)$ smaller than $N_H$.
An Appendix is devoted to a survey
of the Fredholm theory 
in complex space with complex parameters 
[8,9,13,17], which covers the various versions of  
BS-equations encountered in Sec.3 and 4; it  
establishes the corresponding analyticity properties 
and bounds of their solutions in the
relevant variables.

\section{Bethe--Salpeter-type structure of four-point\protect\linebreak %
functions in the $t$-channel: general properties}

We are concerned with relativistic local field theory in 
$(d+1)-$ dimensional space-time, with $d\geq 1$ throughout this section; in the rest of 
the paper, the interesting structures are relevant for all $d \geq 2$. 
As our main object of study, we consider a general four-point function in complex momentum
space $\mathbb{C}^{d+1}$, denoted by $H\left( \left[ k\right] \right) ,$ which describes the
interaction of two local (and mutually local) fields $\phi _{1}$ and $\phi
_{2};$ we specify a certain ``$t$-channel'', with squared-total energy $%
t=\left( k_{1}+k_{2}\right) ^{2}$ as associated with ``two-field states'' of
the form $\left( \int \tilde{\phi}_{1}\left( k_{1}\right) \tilde{\phi}%
_{2}\left( k_{2}\right) f\left( k_{1},k_{2}\right) dk_{1}\;dk_{2}\right)
\left| \Omega \right\rangle,$\  $\left| \Omega \right\rangle $ being the vacuum
state of this field theory and $f$ any appropriate test-function in momentum
space. Following [1], we adapt our notations to this $t-$channel by putting 
$\left[ k\right] =\left( k_{1},k_{2};\ k_{1}^{\prime },k_{2}^{\prime }\right), $
with $k_{1}+k_{2}=k_{1}^{\prime }+k_{2}^{\prime }=K$; 
$k_i$ and $k'_i$ describe incoming and outgoing energy-momentum vectors  
carried by the field $\phi _{i}\left( i=1,2\right) .$ We moreover assume
that the function $H\left( \left[ k\right] \right) $ is {\sl amputated} from the
four external propagator factors\  $\Pi _{i}\left( k_{i}\right) ,\Pi
_{i}\left( k_{i}^{\prime }\right) ,i=1,2,\ \Pi _{i}$ denoting the (complete) two-point
function of the field $\phi _{i}.$ We introduce the incoming and outgoing
relative energy-momentum vectors $Z=\dfrac{k_{1}-k_{2}}{2},\ Z^{\prime }=\dfrac{k_{1}^{\prime
}-k_{2}^{\prime }}{2}$ and we also write $H\left( \left[ k\right] \right)
\equiv H\left( K;Z,Z^{\prime }\right) $ as a function of the three
independent complex vectors $K,Z,Z^{\prime }.$ 

We refer to [1] for a short
survey of the axiomatic analyticity properties of $H\left( \left[ k\right]
\right) .$
A basic fact is that, for any fixed Lorentz frame $LF$ with energy-momentum  
coordinates $K=\left( K^{(0)},\overrightarrow{K}\right) ,Z=\left( Z^{(0)},%
\overrightarrow{Z}\right) ,Z^{\prime }=\left( {Z'}^{(0)},
\overrightarrow{Z^{\prime }}\right) ,$ the function $H\left( \left[ k\right]
\right) $ is holomorphic for all {\sl real or complex} values of the
energy-components $K^{(0)},Z^{(0)},{Z'}^{(0)}$ and {\sl real} values
of the momentum-components 
\break $ \overrightarrow{K},\overrightarrow{Z},%
\overrightarrow{Z^{\prime }} ,$ {\sl except on ``cuts''} which
are defined by {\sl ``spectral sets''} in the various mass and channel-energy
variables 
$$\zeta _{1}=k_{1}^{2}=\left( \dfrac{K}{2} + Z\right) ^{2},\;
\zeta _{2}=k_{2}^{2}=\left( \dfrac{K}{2} - Z\right) ^{2}, $$ 
$$\zeta' _{1}={k'}_{1}^{2}=\left( \dfrac{K}{2} + Z'\right) ^{2},\;
\zeta' _{2}={k'}_{2}^{2}=\left( \dfrac{K}{2} - Z'\right) ^{2}, $$ 
$$ s=\left( k_{1}-k_{1}^{\prime }\right) ^{2}=\left( Z-Z^{\prime
}\right) ^{2},\;u=\left( k_{1}-k_{2}^{\prime }\right) ^{2}=\left(
Z+Z^{\prime }\right) ^{2},\;t=K^{2}.$$ 
In each
of these expressions, the notation $k^{2}$ means ${k^{(0)}}^{2}-\overrightarrow{k}^2$ 
for the corresponding vector $k=\left( k^{(0)},\overrightarrow{k}%
\right) $ varying in $\mathbb{C}^{d+1}$.  
The so-called {\sl ``absorptive parts''} of $H$ are the discontinuities
of $H\left( \left[ k\right] \right) ,$ denoted by $\Delta _{s}H,\;\Delta
_{u}H,\;\Delta _{t}H,$ across the cuts defined respectively by the spectral
sets $\Sigma_{s}\left( s\geqslant s_{0}\right) ,\Sigma_{u}\left( u\geqslant
u_{0}\right) ,$ $\Sigma_{t}\left( t\geqslant t_{0}\right) ;$ $s_{0},t_{0},u_{0}$
denote the corresponding mass thresholds of these spectral sets (here supposed to be 
strictly positive). 
\vskip 0.2cm
It is worthwhile to note that for all {\sl massive} field theories 
{\sl this subset $D_{LF}$ of the primitive axiomatic
analyticity domain contains the whole corresponding ``Euclidean subspace'' }of
complex energy-momentum space, namely the set of all complex configurations $%
\left( K,Z,Z^{\prime }\right) $ whose 
energy-components $(K^{(0)},Z^{(0)},{Z'}^{(0)})$ 
are purely imaginary and whose momentum
components $\left( \overrightarrow{K},\overrightarrow{Z},\overrightarrow{%
Z^{\prime }}\right) $ are real in the Lorentz frame $LF$ considered. 
\vskip 0.2cm

Another aspect which plays an essential role in the following concerns
{\sl ``temperateness properties''} of the four-point function $H\left( \left[ k%
\right] \right) ,$ which we assume to be of the form specified in formula
(3.1) of [1]. These properties imply bounds 
of the following form 
\begin{equation}
\left| H\left( K;Z,Z^{\prime }\right) \right| \leqslant C_{loc}(K,Z,Z')  
(1+ \left\|K\right\|) ^{N_H}(1+\left\| Z\right\|) ^{N_H}(1+\left\| Z^{\prime }\right\|) ^{N_H}, 
\tag{2.1}
\end{equation}
where $N_H$ is a fixed power $\left(N_H \geqslant 0\right) $ and the notation $%
\left\| k\right\| $ stands for the norm of the complex vector $k= (k^{(0)},  \cdots, k^{(d)})$,  
namely $\left\| k\right\| ^{2}= \sum_{0\leq i\leq d} |k^{(i)}|^2;$  
the function $C_{loc}$ includes an inverse power dependence with respect to 
the distance of the point $(K,Z,Z')$ from the union of the spectral sets, 
which takes into account the distribution character of the boundary values of 
$H([k])$ on the reals. 
The bound (2.1) holds uniformly in the 
Euclidean subspace and moreover in the following parts of the 
axiomatic analyticity domain, which are of basic use for the 
BS-structure: 

a) the subset $D_{LF}$  
used in [8,9],  

b) The union of the complex domains $\underline D_{(w,w',\rho')}$ in one vector $\underline k$ 
(for all $w,w',\rho'$) introduced in
[1] (see Propositions 3 and 4 of the latter). 

\subsection{A recall of axiomatic results on the Bethe-Salpeter structure of
four-point functions}

In [8,9], it has been shown that for massive QFT's satisfying the
postulate of \textit{``asymptotic completeness of two-particle states''} (or
``off-shell unitarity in the two particle spectral region'') and an 
additional regularity assumption in the energy-variable $t$ (for $t \geq t_0$), the
two-particle $t$-channel analytic structure of $H\left( \left[ k\right]
\right) $ is entirely encoded in any Bethe--Salpeter-type integral equation
of the following form 
\begin{equation}
H\left( K;Z,Z^{\prime }\right) =B\left( K;Z,Z^{\prime }\right) +\int_{\Gamma
\left( K\right) }
B\left( K;Z,Z^{\prime \prime }\right) 
H\left( K;Z^{\prime \prime },Z^{\prime }\right) G\left(
K;Z^{\prime \prime }\right)\ dZ^{\prime \prime }  \tag{2.2}
\end{equation}%
where the ``Bethe--Salpeter kernel'' $B\left( \left[ k\right] \right) \equiv
B\left( K;Z,Z^{\prime }\right) $ is a four-point function satisfying the
same axiomatic analyticity properties and bounds of the form (2.1)  
\footnote{%
These analyticity properties and bounds of $B$ are implied by those of $H,$ except for
possible ``CDD poles'' in $K$ corresponding to the Fredholm alternative,
which can be excluded for $t$ negative and $|t|$ sufficiently large
under a suitable choice of the regularized
propagators $\Pi_i^{(reg)}$ (see Appendix, proposition A1).}  
as $H\left( \left[ k\right] \right) $
and in addition the property of {\sl ``two-particle irreducibility in the $t$%
-channel''.} This means that the corresponding absorptive part (or
discontinuity) $\Delta_t B$ of $B$ vanishes in the two-particle region, 
which is of the form $\left\{ t;t_{0}=\left( m_{1}+m_{2}\right) ^{2}\leqslant t <  
M^{2}\right\} ,m_{1}$ and $m_{2}$ being the masses of the
``lowest poles'' $\left( k_{i}^{2}=m_{i}^{2},\ i=1,2\right) $ in the respective
propagators $\Pi_{1}\left( k_{1}\right) ,\Pi_{2}\left( k_{2}\right) $ of 
$\phi _{1}$ and $\phi _{2}.$

As a matter of fact, these poles are present in the function $G\left(
K,Z^{\prime \prime }\right) $ under the integral of Eq.(2.2), which
represents a ``regularized double-propagator'' with respect to the internal $(d+1)-$momenta $%
k_{i}^{\prime \prime }=\dfrac{K}{2}\pm Z^{\prime \prime },i=1,2,$ of the
following form: 
\begin{equation}
G\left( K,Z^{\prime \prime }\right) =i\Pi_{1}^{\left( \mathrm{reg}%
\right) }\left( \dfrac{K}{2}+Z^{\prime \prime }\right)
.\Pi_{2}^{\left( \mathrm{reg}\right) }\left( \dfrac{K}{2}%
-Z^{\prime \prime }\right) ;  \tag{2.3}
\end{equation}
$\Pi_{i}^{\left( \mathrm{reg}\right) } \ (i=1,2)$ is a regularized form of $%
\Pi_{i}$ obtained by multiplying the latter by a suitable
Pauli--Villars-type factor, equal to 1 on the mass shell. The only required
property is that $\Pi_{i}^{\left( \mathrm{reg}\right)}$  
be a Lorentz-invariant function, namely a function of the 
squared-mass complex variable $\zeta_i = k_i^2$ having the same pole 
$\zeta_i = m_i^2$  
(with the same residue) as $\Pi_i$, analytic in the same cut-plane of the form  
$\mathbb{C}\backslash \{ m_{i}^{2}\} \backslash [M_i^2, +\infty [ $ and  
satisfying uniform bounds of the following form 
\begin{equation}
\left| \Pi_{i}^{\left( \mathrm{reg}%
\right) }\left( k_{i}\right) \right| \leqslant \left[c(1+\left\|
k_{i}\right\|^2)\right] ^{-r}, \tag{2.4}  
\end{equation}
with $r$ sufficiently large
and $c <1$, for $%
k_{i}$ Euclidean (i.e. $\zeta _{i}=-\left\| k_{i}\right\| ^{2}$).  

The integration cycle $\Gamma \left( K\right) $ in Eq. (2.2) is the
Euclidean subspace $E_{d+1}=\left\{ Z^{\prime \prime }=\left( i{Y''}^{(0)},   
\overrightarrow{X}^{\prime \prime }\right) \right\} ,$ when $K,Z$
and $Z^{\prime }$ are themselves taken in $E_{d+1}.$

If $r$ is sufficiently large with respect to $N,$ Eq.
(2.2) is a genuine Fredholm integral equation (depending analytically on the
vector parameter $K$) which allows one to define $B\left( \left[ k\right]
\right) $ in terms of $H\left( \left[ k\right] \right) ,$ with $B$ also
satisfying bounds of the form (2.1). In [8,9], Eq. (2.2) has
been shown to extend by analytic continuation to all configurations  $\left(
K,Z,Z^{\prime }\right) $ in the subset $D_{LF}$ of the axiomatic domain,  
provided $\Gamma \left( K\right) $ is distorted from its initial situation $%
E_{d+1}= E_{d+1}(LF)$ in order to remain in the analyticity domain of the integrand and
to have an infinite part parallel to $E_{d+1}.$ 
(Of course, by using the Lorentz invariance of the axiomatic domain, Eq.(2.2) is also 
shown to sweep similarly the subsets $D_{LF}$ associated with all Lorentz frames $LF$).
In the following, we shall
call {\sl ``Feynman convolution''} and denote by $\left( B\,\circ_{t} H\right) \left(
K;Z,Z^{\prime }\right)$  
the integral at the r.h.s. of Eq.(2.2), which enjoys the 
axiomatic analyticity properties of a four-point function. Eq.(2.2) can thus be
rewritten in short: 
\begin{equation}
H=B+B\,\circ_{t}H,  \tag{2.2'}
\end{equation}
The analysis of [8,9] relies on the exploitation of Eq. (2.2) as a
Fredholm resolvent integral equation on the space $\Gamma \left( K\right) ,$
depending analytically of the (vector) parameter $K.$ In particular, the
result of this analysis is that $H\left( \left[ k\right] \right) $ can be
written as a ratio of the following form 
\begin{equation}
H\left( K;Z,Z^{\prime }\right) =\dfrac{{\cal N}_B\left( K;Z,Z^{\prime }\right) }{%
{\cal D}_B\left( K^{2}\right) }  \tag{2.5}
\end{equation}%
where ${\cal N}_B\left( K;Z,Z^{\prime }\right) $ is a four-point function and $%
{\cal D}_B\left( K^{2}\right) $ is a two-point function, defined in terms of $B$ and $%
G$ through the appropriate Fredholm series. ${\cal N}_B$ and ${\cal D}_B$ are proven to admit
analytic continuations in the variable $t=K^{2}$ across the two-particle
region $t_{0}\leqslant t<M^{2}$ in a ramified domain around the two-particle
threshold $t=t_{0}.$ The zeros of ${\cal D}_B \left( K^{2}\right) $ correspond to
poles of $H\left( \left[ k\right] \right) ,$ interpreted as bound states or
resonances.

\subsection{Bethe-Salpeter equations for the absorptive parts in the crossed channels}

In the present paper, we shall not exploit Eq. (2.2) for the latter
meromorphic structure in $t$ occurring at $\mathrm{\func{Re}\,}t>0,$ but for
its remarkable implications in the crossed channels, namely the $s$ and $u$%
-channels.\ In fact, it has been shown in [8,10] that Eq. (2.2) implies
corresponding integral relations between the absorptive parts $\Delta
_{s}H,\Delta _{u}H$ of $H$ and $\Delta _{s}B,\Delta _{u}B$ of $B,$ which
need to be recalled with some care.\ In these relations, the Feynman
convolution $\circ_{t}$ is replaced by a \textit{composition product} $%
\diamondsuit_{t}$ involving integration on a certain \textit{compact} set
specified below. We now write these relations in the following concise form,
before giving their detailed meaning and interpretation: 
\begin{equation}
\Delta _{s}H=\Delta _{s}B+\Delta _{s}B\,\diamondsuit _{t}\,\Delta
_{s}H+\Delta _{u}B\,\diamondsuit _{t}\,\Delta _{u}H,  \tag{2.6}
\end{equation}%
\begin{equation}
\Delta _{u}H=\Delta _{u}B+\Delta _{s}B\,\diamondsuit _{t}\,\Delta
_{u}H+\Delta _{u}B\,\diamondsuit _{t}\,\Delta _{s}H;  \tag{2.7}
\end{equation}

In the latter, the absorptive part $\Delta _{s}H$ (and similarly for $\Delta
_{s}B)$ is specified as follows: 
\begin{equation}
\begin{array}{lllll}
\!\Delta _{s}H\!\!\left( \!K\!;\!Z\!,\!Z^{\prime }\right) \!= i & \!\!\left[
\lim \!\right. \!\! & H\!\left( \!K\!;\!Z\!,\!Z^{\prime }\right)\  \!-\!\!
& \!\!\lim & \!\!\left. H\!\left( \!K\!;\!Z\!,\!Z^{\prime }\right) 
\right] \\ 
& \!\!\func{Im}\left( \!Z\!\!-\!\!Z^{\prime }\right) ^{2}\!\!=\!-\!\varepsilon &  & 
\!\!\func{Im}\left( \!Z\!\!-\!\!Z^{\prime }\right) ^{2}\!\!=\!\!\varepsilon
&  \\ 
& \varepsilon >0\quad \varepsilon \rightarrow 0 &  & \varepsilon >0\quad
\varepsilon \rightarrow 0 & 
\end{array}
\tag{2.8}
\end{equation}
The support of $\Delta _{s}H$ (or $\Delta _{s}B),$ which is the
spectral set 
\begin{equation*}
\Sigma_{s}\ = \{(K,Z,Z');\ k_{1}-k_{1}^{\prime }=Z-Z^{\prime }\func{real},\ \ \ \left(
Z-Z^{\prime }\right)^2 \geq s_{0}\}  
\end{equation*}
is contained in the union of the following two disjoint sets 
\begin{equation*}
\Sigma_{s}^{+}=\left\{ \left( K;Z,Z^{\prime }\right) ;\ Z-Z^{\prime }%
\func{real},\ Z-Z^{\prime }\in 
V^{+}\right\},  
\end{equation*}
\begin{equation*}
\Sigma_{s}^{-}=\left\{ \left( K;Z,Z^{\prime }\right) ;Z-Z^{\prime }%
\func{real},\ Z-Z^{\prime }\in  
V^{-}\right\}, 
\end{equation*}
where $V^+$ (resp. $V^-$) denotes the open forward (resp. backward) cone: \break 
$k^2 \equiv {k^{(0)}}^2 - {\vec k}^2 >0,\  k^{(0)} >0$ (resp. $k^{(0)} <0$).   

We shall call $\Delta _{s}^{+}H$ and $\Delta _{s}^{-}H$ (resp. $\Delta
_{s}^{+}B$ and $\Delta _{s}^{-}B)$ the restrictions of $\Delta _{s}H$ (resp. 
$\Delta _{s}B)$ to the corresponding disjoint sets $\Sigma_{s}^{+}$
and $\Sigma_{s}^{-}.$

\vskip 0.3cm 
The absorptive part $\Delta _{u}H$ (and similarly for $\Delta _{u}B)$ is
specified as follows:%
\vskip 0.3cm%

\begin{equation}
\begin{array}{lllll}
\!\Delta _{u}H\!\!\left( \!K\!;\!Z\!,\!Z^{\prime }\right) \!= i & \!\!\left[
\lim \!\right. \!\! & \!\!H\!\left( \!K\!;\!Z\!,\!Z^{\prime }\right)\  \!-\!\!
& \!\!\lim & \!\!\left. \!\!H\!\left( \!K\!;\!Z\!,\!Z^{\prime }\right) 
\right] \\ 
& \!\!\func{Im}\left( \!Z\!\!+\!\!Z^{\prime }\right) ^{2}\!\!=\!-\!\varepsilon &  & 
\!\!\func{Im}\left( \!Z\!\!+\!\!Z^{\prime }\right) ^{2}\!\!=\!\!\varepsilon
&  \\ 
& \varepsilon >0\quad \varepsilon \rightarrow 0 &  & \varepsilon >0\quad
\varepsilon \rightarrow 0 & 
\end{array}
\tag{2.9}
\end{equation}
\vskip 0.3cm%

The support of $\Delta _{u}H$ (or $\Delta _{u}B),$ which is the
spectral set 
\begin{equation*}
\Sigma_{u}\ =\ \{(K,Z,Z');\  k_{1}-k_{2}^{\prime }=Z+Z'\ \func{real},\ \left(
Z+Z^{\prime }\right) ^{2}\geq u_{0}\} ,
\end{equation*}
is contained in the union of the following two disjoint sets 
\begin{equation*}
\Sigma_{u}^{+}=\left\{ \left( K;Z,Z^{\prime }\right) ;\ Z+Z^{\prime }%
\func{real},\ Z+Z^{\prime }\in V^{-}\right\},  
\end{equation*}
\begin{equation*}
\Sigma_{u}^{-}=\left\{ \left( K;Z,Z^{\prime }\right) ;\ Z+Z^{\prime }%
\func{real},\ Z+Z^{\prime }\in V^{+}\right\} .
\end{equation*}

The corresponding restrictions of $\Delta _{u}H\left( \mathrm{resp.}\Delta
_{u}B\right) $ to $\Sigma_{u}^{+}$ and $\Sigma_{u}^{-}$ are
called $\Delta _{u}^{+}H$ and $\Delta _{u}^{-}H\left( \mathrm{resp.}\Delta
_{u}^{+}B\mathrm{\ and\ }\Delta _{u}^{-}B\right) .$

\vskip 0.3cm 
For all complex configurations $\left( K;Z,Z^{\prime }\right) $ in $%
\Sigma_{s}^{+}$ one introduces the ``double-cone''  
\begin{equation*}
\diamondsuit \left( Z,Z^{\prime }\right) =\left\{ Z^{\prime \prime
};\ Z-Z^{\prime \prime }\ \func{real},\ Z-Z^{\prime \prime }\in V^{+},\ Z^{\prime
\prime }-Z^{\prime }\in V^{+}\right\}
\end{equation*}
and defines the composition product $\diamondsuit _{t}$ as follows: 
\begin{equation*}
\left( \Delta _{s}B\ \diamondsuit _{t}\ \Delta _{s}H\right) \left( K;Z,Z^{\prime
}\right) =
\end{equation*}
\vspace{-0.55cm} 
\begin{equation}
\int_{\diamondsuit \left( Z,Z^{\prime }\right) }\Delta _{s}^{+}B\left(
K;Z,Z^{\prime \prime }\right) \Delta _{s}^{+}H\left( K;Z^{\prime \prime
},Z^{\prime }\right) G\left( K;Z^{\prime \prime }\right) dZ^{\prime \prime }
\tag{2.10}
\end{equation}
and similarly 
\begin{equation*}
\left( \Delta _{u}B\ \diamondsuit _{t}\ \Delta _{u}H\right) \left( K;Z,Z^{\prime
}\right) =
\end{equation*}
\vspace{-0.55cm} 
\begin{equation}
\int_{\diamondsuit \left( -Z^{\prime },-Z\right) }\Delta _{u}^{-}B\left(
K;Z,Z^{\prime \prime }\right) \Delta _{u}^{+}H\left( K;Z^{\prime \prime
},Z^{\prime }\right) G\left( K;Z^{\prime \prime }\right) dZ^{\prime \prime }
\tag{2.11}
\end{equation}

It is important to note (in view of the distribution character of the absorptive
parts $\Delta _{s,u}H$ and $\Delta _{s,u}B$ in the corresponding variables 
$s$ and $u$ respectively) that these
composition products are always well-defined in the sense of distributions
in view of the compactness of the integration region.

Eqs. (2.10) and (2.11) specify completely the meaning of formula (2.6) in the
case when $Z-Z^{\prime }\in V^{+}.$

The complete specification of formulae (2.6) and (2.7) for configurations $%
\left( K,Z,Z^{\prime }\right) $ contained respectively in $%
\Sigma_{s}^{-},\Sigma_{u}^{+}$ and $\Sigma_{u}^{-}$ is  
obtained by changing the integration region $\diamondsuit \left( Z,Z^{\prime
}\right) $ or $\diamondsuit \left(-Z',-Z\right) $ of (2.10), (2.11)
into an appropriate region of the form 
$\diamondsuit \left( \varepsilon
Z,\varepsilon ^{\prime }Z^{\prime }\right)$ 
or $\diamondsuit \left( \varepsilon' 
Z',\varepsilon Z\right) ,$ 
with $\varepsilon ,\varepsilon
^{\prime }=\pm 1$ and by picking-up the corresponding relevant parts $\Delta
_{s,u}^{\pm }H$ and $\Delta _{s,u}^{\pm }B$ whose choice is dictated by the
consistency of support properties.

The derivation of formulae (2.6), (2.7), which is obtained by taking the
absorptive parts $\Delta _{s}$ and $\Delta _{u}$ of both sides of Eq. (2.2),
relies on the following discontinuity formulae;  
for any pair of four-point functions $%
\left( F_{1},F_{2}\right) ,$ one has: 
\begin{equation}
\Delta _{s}\left( F_{1}\circ_{t}F_{2}\right) =\Delta _{s}F_{1}\ \diamondsuit
_{t}\ \Delta _{s}F_{2}+\Delta _{u}F_{1}\ \diamondsuit _{t}\ \Delta _{u}F_{2}
\tag{2.12}
\end{equation}
and 
\begin{equation}
\Delta _{u}\left( F_{1}\circ_{t}F_{2}\right) =\Delta _{s}F_{1}\ \diamondsuit
_{t}\ \Delta _{u}F_{2}+\Delta _{u}F_{1}\ \diamondsuit _{t}\ \Delta _{s}F_{2} 
\tag{2.13}
\end{equation}
These formulae have been derived in [8,10] by performing suitable 
distortions of the integration cycle $\Gamma(K)$ (starting from 
$E_{d+1}$) in the 
Feynman-convolution product
$(F_1 \circ_t F_2)(K;Z,Z') = \int_{\Gamma(K)} F_1(K;Z,Z'') F_2(K;Z'',Z') 
G(K;Z'') dZ''.$   
These distortions, which are performed in the energy variable ${Z''}^{(0)}$  
inside the axiomatic domain of the integrand, lead one to fold the cycle 
$\Gamma(K)$ around the supports of the absorptive parts $\Delta_{s,u} F_1$ 
and $\Delta_{s,u} F_2$ in limiting situations  
$\Gamma(K) = \Gamma_{s\pm}(K) $ (resp. $\Gamma_{u\pm}(K)$)  
where $s = \Re e s \pm i\eta$ 
(resp.  $u = \Re e u \pm i\eta$), with $\eta$ positive and tending to zero.
Taking the discontinuities $\Delta_s$
(resp. $\Delta_u$) between the convolution integrals on   
$ \Gamma_{s +}(K) $ 
and $ \Gamma_{s -}(K) $ 
(resp. $\Gamma_{u+}(K)$  
and $ \Gamma_{u -}(K) $) then reduces the integration cycle to the 
compact cycle with support 
$\diamondsuit \left( \varepsilon
Z,\varepsilon ^{\prime }Z^{\prime }\right)$ 
previously defined.

A nice property enjoyed by the previous formulae (2.12), (2.13) is the following 

\vskip 0.3cm
\noindent
{\bf Additivity property of spectral sets:}

\vskip 0.3cm
\noindent
If two four-point functions $F_{1}\left( \left[ k\right] \right)
,F_{2}\left( \left[ k\right] \right) $ have their spectral sets $\Sigma_{s}$
and $\Sigma_{u}$ specified respectively by the conditions $s\geq s_{1},\ u\geq
u_{1}$ and $s\geq s_{2},\  u\geq u_{2},$ then the following support
properties hold, \textit{inside each connected component} $%
\Sigma_{s}^{+},\Sigma_{s}^{-},\Sigma_{u}^{+},\Sigma_{u}^{-}$ of the spectral sets:%
\vskip 0.3cm%

a) The support of 
$\left( \Delta _{s}F_{1}\ \diamondsuit _{t}\ \Delta_s F_{2}\right) \left( K,Z,Z^{\prime
}\right)$  
is contained in the set 
\begin{equation*}
\left\{ \left( K,Z,Z^{\prime }\right) ;\ \left( Z-Z^{\prime }\right) ^{2}\geq
\left( \sqrt{s_{1}}+\sqrt{s_{2}}\right) ^{2}\right\} .
\end{equation*}

This directly follows from the definition (2.10) of $\diamondsuit _{t}$ for $%
\Delta _{s}F_{1}\ \diamondsuit _{t}\ \Delta _{s}F_{2}$ by using the fact that $%
Z-Z^{\prime }=\left( Z-Z^{\prime \prime }\right) +\left( Z^{\prime \prime
}-Z^{\prime }\right) ,$ with $\left( Z-Z^{\prime \prime }\right) ^{2}\geq
s_{1}$ and $\left( Z^{\prime \prime }-Z^{\prime }\right) ^{2}\geq s_{2}$ $%
\left( Z-Z^{\prime }\mathrm{\ }\text{can be either in }V^{+}\text{ or in }%
V^{-}\right). $%

b) Similarly, the supports of $\Delta _{u}F_{1}\ \diamondsuit _{t}\ \Delta
_{u}F_{2},\ \Delta _{s}F_{1}\ \diamondsuit _{t}\ \Delta _{u}F_{2},\ \Delta
_{u}F_{1}\ \diamondsuit _{t}\ \Delta _{s}F_{2}$ are respectively contained in
the sets defined by the conditions: 
\begin{eqnarray*}
\left( Z-Z^{\prime }\right) ^{2} &\geq &\left( \sqrt{u_{1}}+\sqrt{u_{2}}%
\right) ^{2},\\ 
\left( Z+Z^{\prime }\right) ^{2} 
&\geq &\left( \sqrt{s_{1}}+\sqrt{u_{2}}\right) ^{2},\\  
\left( Z+Z^{\prime
}\right) ^{2} 
&\geq &\left( \sqrt{u_{1}}+\sqrt{s_{2}}\right) ^{2}
\end{eqnarray*}
(with all the possibilities $Z-Z^{\prime }\in V^{\pm }$ and $Z+Z' \in V^{\pm}$).  

\vskip 0.3cm 
Formulae (2.6) and (2.7) appear as a pair of coupled Fredholm-type
equations (depending on $K)$ which we call \textit{``Bethe--Salpeter
equations for the crossed-channel absorptive-parts''.}
A remarkable feature of these equations, which is due to the ``additivity
property of spectral sets'', is the following

\vskip 0.3cm
\noindent
{\bf Finiteness property of the Bethe-Salpeter equations for 
crossed-channel absorptive-parts} (theorem 1 of [8]): 

\vskip 0.3cm 
{\sl In any bounded region of the
variables $s$ and $u,$ the Bethe-Salpeter equations (2.6) and (2.7) for 
$\Delta_s H$ and $\Delta_u H$ can be solved explicitly by a {\rm finite }%
number of $\diamondsuit _{t}$-composition-products}.

\vskip 0.2cm 
In fact, by applying the standard iteration procedure to Eq. (2.6), one gets
the following relations, written for simplicity in the case when $%
s_{0}=u_{0}:$

\vskip 0.3cm 
For 
$s<4s_{0},\qquad  \Delta _{s}H=\Delta _{s}B.$ 

\vskip 0.3cm 
For 
$s<9s_{0}, \qquad \Delta _{s}H=\Delta _{s}B+\Delta _{s}B\ \diamondsuit
_{t}\ \Delta _{s}B\ +\ \Delta _{u}B\ \diamondsuit _{t}\ \Delta _{u}B.$ 

\vskip 0.3cm 
For 
$s<16s_{0},$  
\begin{equation*}
\Delta _{s}H=\Delta _{s}B+\Delta _{s}B \ \diamondsuit _{t}\ \Delta _{s}B+\Delta
_{u}B\ \diamondsuit _{t}\ \Delta _{u}B+\Delta _{s}B \ \diamondsuit _{t}\ \Delta
_{s}B\ \diamondsuit _{t}\ \Delta _{s}B
\end{equation*}
\vspace{-0.55cm} 
\begin{equation}
+\Delta _{s}B\ \diamondsuit _{t}\ \Delta _{u}B\ \diamondsuit _{t}\ \Delta
_{u}B+\Delta _{u}B\ \diamondsuit _{t}\ \Delta _{s}B\ \diamondsuit _{t}\ \Delta
_{u}B+\Delta _{u}B\ \diamondsuit _{t}\ \Delta _{u}B\ \diamondsuit _{t}\ \Delta _{s}B
.\tag{2.14}
\end{equation}
etc...

\vskip 0.3cm 
Eq. (2.7) is solved by similar expressions for $\Delta _{u}H.$

\section{Transferring the Bethe-Salpeter structure  
from complex momentum space to mass and complex-angular-momentum variables} 

In this section, we shall derive alternative versions of the 
Bethe-Salpeter equations (2.2) and  (2.6),(2.7) in which the complex energy-momentum 
vectors are replaced 1) by {\sl squared-mass and complex-angular variables} 
and 2) by {\sl squared-mass and complex-angular-momentum (CAM) variables}.
Throughout these two steps, the four-point function $ H([k])$ 
is considered as given with 
its axiomatic analyticity properties and temperate bounds of the form (2.1), 
which imply (according to the results of [1])  
two corresponding sets of properties in the complex-angular and CAM variables. 
At each step, analogous properties of the Bethe-Salpeter kernel 
$B$ will appear as derived from those of $H$ 
by applying the Fredholm method or {\sl ``$\cal N / \cal D$-method''} in complex-space 
to the corresponding version of the Bethe-Salpeter equation. 
The common point to these various versions will be the occurrence of an integration-space
involving radial and longitudinal variables $(\rho,w)$, equivalent to
``squared-mass variables'' $\zeta\equiv (\zeta_1,\zeta_2),$ as described below.
A unified presentation of the relevant results of the  
$\cal N / \cal D$-method in complex-space, 
applicable to these various versions  
is summarized in the Appendix .

\vskip 0.3cm
We now assume that $d\geq 2$ and 
fix once for all the total energy-momentum vector 
$K$ of the $t$-channel in such a way that $K$ is space-like (i.e. $%
t<0 $); we choose a system of space-time coordinates such
that $K=\left( 0,0,...,0,\sqrt{-t}\right) .$
For real or complex vectors $k=(k^{(0)},k^{(1)},\ldots,k^{(d)}),\  
k'=({k'}^{(0)},{k'}^{(1)},\ldots,{k'}^{(d)}),$ the Minkowskian scalar product  
is $k\cdot k' =k^{(0)} {k'}^{(0)} -k^{(1)}{k'}^{(1)}- \cdots k^{(d)} {k'}^{(d)}$ 
and $k^2\equiv k\cdot k$.  

We adopt the parametrization of the vector variables $Z,Z^{\prime }$ in
terms of \textit{radial, longitudinal and angular variables} $\rho ,w,z$ and 
$\rho ^{\prime },w^{\prime },z^{\prime }$ as in [1], Eq.
(2.2), namely 
\begin{equation}
Z=\rho z+w K\quad ,\quad Z^{\prime }=\rho ^{\prime }z^{\prime }+w^{\prime }K,
\tag{3.1}
\end{equation}
where the vectors $z=\left( z^{(0)},z^{(1)},...,z^{(d-1)},0\right),\  
z'=\left({z'}^{(0)}, {z'}^{(1)},...,{z'}^{(d-1)} ,0\right) $
are such that: 
\begin{equation}
z\cdot K=z^{\prime }\cdot K=0\qquad \mathrm{and}\qquad z^{2}=z^{\prime 2}=-1. 
\tag{3.2}
\end{equation}

The radial and longitudinal variables $\left( \rho ,w\right) $ (resp. $%
\left( \rho ^{\prime },w^{\prime }\right) $) can be equivalently replaced by
the ``\textit{squared-mass variables'' }$\zeta =\left( \zeta _{1},\zeta
_{2}\right) ,\mathrm{\ resp.\ }\zeta ^{\prime }=\left( \zeta _{1}^{\prime
},\zeta _{2}^{\prime }\right) ,$ where 
\begin{equation}
\zeta _{1,2} =\left( \dfrac{K}{2}\pm Z\right) ^{2} = -\rho^2 + (w\pm {1\over 2})^2 t, \ 
\zeta' _{1,2} =\left( \dfrac{K}{2}\pm Z'\right) ^{2} = -{\rho'}^2 + (w'\pm {1\over 2})^2 t 
\ \ \tag{3.3} 
\end{equation}
and one has:
\begin{equation}
\ \rho ^{2} =\dfrac{\Lambda \left( \zeta _{1},\zeta _{2},t\right) }{4t},
\qquad \qquad\ \ \ \rho
^{\prime 2}=\dfrac{\Lambda \left( \zeta _{1}^{\prime },\zeta _{2}^{\prime
},t\right) }{4t},  
\tag{3.4} 
\end{equation}
with 
\begin{equation}
\Lambda \left( \alpha ,\beta ,\gamma \right) =\left( \alpha -\beta
\right)^2 -2\left( \alpha +\beta \right) \gamma +\gamma ^{2}  \tag{3.5}
\end{equation}
and 
\begin{equation}
w=\dfrac{\zeta _{1}-\zeta _{2}}{2t}\quad ,\quad w^{\prime }=\dfrac{\zeta
_{1}^{\prime }-\zeta _{2}^{\prime }}{2t}.  \tag{3.6}
\end{equation}

The variables $\rho ,w,\rho ^{\prime },w^{\prime }$ will always be real, with $%
\rho >0,\rho ^{\prime }>0,$ which means that  $(\zeta ,\zeta
^{\prime })$ varies in the real region $\Delta _{t}\times \Delta _{t},$ where $%
\Delta _{t}$ is the parabolic region of (negative) ``Euclidean
squared-masses'' 
(see fig.\ 1 of [1])
\begin{equation*}
\Delta _{t}=\left\{ \zeta =\left( \zeta _{1},\zeta _{2}\right) ;\Lambda
\left( \zeta _{1},\zeta _{2},t\right) <0\right\}.  
\end{equation*}%

On the contrary, the vector variables $z,z^{\prime }$ 
or {\sl ``hyperbolic angular variables''} 
are allowed to
vary on the whole \textit{complex} hyperboloid $X_{d-1}^{\left( c\right) }$
with equation $z^{2}=z^{\prime 2}=-1$ (in the subspace $z^{(d)}= {z'}^{(d)} =0$) 
and one introduces the variable $\cos \Theta _{t}=-z.z^{\prime
},\ \Theta _{t}$ being interpreted as the off-shell (complex) scattering angle
in the $t$-channel.

\subsection{Bethe--Salpeter equation in the mass variables and\allowbreak
angular variables - Partial-wave decomposition in the Euclidean region}

In this subsection, we shall assume that the vectors $Z,Z^{\prime }$ remain 
\textit{in the Euclidean subspace} $E_{d+1}$ which in the parametrization
(3.1), (3.2) corresponds to choosing the complex vectors $z,z^{\prime }$ on
a unit sphere of dimension $d-1,$ called ``the Euclidean sphere'' $S_{d-1}$ of $X_{d-1}^{(c)}$, 
namely
\begin{equation*}
z=\left( iy^{(0)},x^{(1)},...,x^{(d-1)},0\right) ,\ z' =\left(
i{y'}^{(0)}, {x'}^{(1)},..., {x'}^{(d-1)},0 \right) ,
\end{equation*} 
with 
\begin{equation*}
\omega =\left( y^{(0)},x^{(1)},...,x^{(d-1)}\right) \in \mathbb{S}_{d-1}, \ \  
\omega' =\left(
{y'}^{(0)}, {x'}^{(1)},..., {x'}^{(d-1)} \right) 
\in \mathbb{S}_{d-1}. 
\end{equation*}
The Minkowskian scalar product $z\cdot z'$ here reduces to 
$z\cdot z' = -\left\langle \omega \cdot \omega ^{\prime }\right\rangle ,$
where $\left\langle \omega \cdot \omega ^{\prime
}\right\rangle $ denotes the Euclidean scalar product in $\mathbb{R}^{d}.$
Then $\left\langle \omega \cdot \omega ^{\prime }\right\rangle =\cos \Theta _{t}$
with $\Theta _{t}$ real.  

\vskip 0.2cm
We then intend to write the Bethe--Salpeter equation (2.2) in the Euclidean
subspace $E_{d+1}$ in terms of the radial, longitudinal and angular
variables. We shall use the fact that the four-point function $H\left( \left[
k\right] \right) $ of the scalar fields $\left( \phi _{1},\phi _{2}\right) ,$
and therefore also $B\left( \left[ k\right] \right) $ are invariant under
the connected part of the complex Lorentz group $SO_{0}^{\left( c\right)
}\left( 1,d\right) .$ This implies that these functions only depend on the
Lorentz-invariant variables $\left( \rho ,w ,\rho ^{\prime }, w 
^{\prime },t\right) ,$ or equivalently $\left( \zeta ,\zeta ^{\prime
},t\right) $ and 
$z\cdot z^{\prime }=-\left\langle \omega \cdot \omega ^{\prime
}\right\rangle =-\cos \Theta _{t},$
the restrictions of these functions 
to the Euclidean subspace $(E_{d+1})^3$ 
being then invariant under the connected orthogonal group $%
SO_{0}\left( d+1\right) $. One can thus put 
(with notations similar to those of 
[1], Eq. (3.24)):
\begin{equation}
H\left( \left[ k\right] \right) \equiv H\left[ t;\rho ,w ;\rho ^{\prime
},w^{\prime };\left\langle \omega \cdot \omega ^{\prime
}\right\rangle \right] \equiv \underline{H}_{\left( \zeta ,\zeta ^{\prime
},t\right) }\left( \left\langle \omega \cdot \omega ^{\prime
}\right\rangle \right)  \tag{3.7}
\end{equation}
\begin{equation}
B\left( \left[ k\right] \right) \equiv B\left[ t;\rho ,w ;\rho ^{\prime
},w^{\prime };\left\langle \omega \cdot \omega ^{\prime
}\right\rangle \right] \equiv \underline{B}_{\left( \zeta ,\zeta ^{\prime
},t\right) }\left( \left\langle \omega \cdot \omega ^{\prime
}\right\rangle \right)  \tag{3.8}
\end{equation}

We note that in view of Eq(3.1),
the bounds (2.1) on $H$ in 
$(E_{d+1})^3$ 
can be rewritten as follows , with a suitable constant $C_E^{(H)}:$ 
$$| H\left[ t;\rho ,w ;\rho ^{\prime },w^{\prime} 
;\left\langle \omega \cdot \omega ^{\prime }\right\rangle \right] |\  \leq $$ 
\begin{equation}
C_E^{(H)}  (1+|t|^{1\over 2})^{N_H} \ (1+\rho)^{N_H}\ (1+ |w||t|^{1\over 2})^{N_H} 
 \ (1+\rho')^{N_H}\ (1+ |w'||t|^{1\over 2})^{N_H}.  
\tag{3.9}    
\end{equation}

In view of the Lorentz invariance of the propagator 
$\Pi_{i}^{\left( \mathrm{reg}\right) }\left( k\right) \equiv \Pi_{i}^{\left( 
\mathrm{reg}\right) }\left[ k^{2}\right] $ of $\phi _{i}\left( i=1,2\right),$  
Eq. (2.3) can also be rewritten as follows: 
\begin{equation}
G\left( K,Z\right) =\underline{G}\left( \zeta \right)
=i\Pi_1^{\left( \mathrm{reg}\right) }\left[ \zeta _{1}\right]
\Pi_2^{\left( \mathrm{reg}\right) }\left[ \zeta _{2}\right]
\tag{3.10}
\end{equation}
or 
\begin{equation}
G\left[ t;\rho ,w\right] =i\Pi_1^{\left( \mathrm{reg}\right) }\left[
\left( w+ 1/2 \right) ^{2}t-\rho^{2}\right] \Pi_2^{\left( 
\mathrm{reg}\right) }\left[ \left( w-1/2\right) ^{2}t-\rho^{2}\right],  
\tag{3.11}
\end{equation}
and the bounds (2.4) then yield correspondingly:
\begin{equation}
|G[t;\rho,w]|\leq c^2  
[1+\rho^2 +|t|(w+ {1\over 2})^2]^{-r} 
[1+\rho^2 +|t|(w- {1\over 2})^2]^{-r}.  
\tag{3.12}
\end{equation}
In view of (3.1) ...(3.6), 
the integration measure $dZ$ on $E_{d+1}$ reads: 
\begin{equation}
dZ=i\sqrt{-t} \rho^{d-1}d\rho \,dw\,d\omega =-id\mu _{t}\left( \zeta \right)
d\omega  \tag{3.13}
\end{equation}
with 
\begin{equation}
d\mu _{t}\left( \zeta \right) =\dfrac{1}{4\sqrt{-t}}\left( \dfrac{\Lambda\left(
\zeta _{1},\zeta _{2},t\right) }{4t}\right) ^{\dfrac{d-2}{2}}d\zeta
_{1,}d\zeta _{2}  \tag{3.14}
\end{equation}
We can now give the following alternative form of the BS-equation (2.2) 
in Euclidean space in terms of mass variables and angular variables; 
as shown in the Appendix, the bounds (3.9), (3.12) on $H$ and $G$ 
ensure that this integral relation is a genuine Fredholm equation 
depending on the parameter $t$ for all $t <0$. One obtains: 
\begin{equation*}
H\left[ t;\rho ,w;\rho ^{\prime },w^{\prime };\left\langle \omega \cdot 
\omega ^{\prime }\right\rangle \right] =B\left[ t;\rho ,w;\rho ^{\prime
},w^{\prime };\left\langle \omega \cdot  \omega ^{\prime }\right\rangle %
\right]
\end{equation*}
\vspace{-0.85cm} 
\begin{equation}
+\left( B{\circ_{t}}H\right) \left[ t;\rho ,w;\rho ^{\prime },w^{\prime
};\left\langle \omega \cdot \omega ^{\prime }\right\rangle \right] , 
\tag{3.15}
\end{equation}
where: 
\begin{equation*}
\left( B{\circ_{t}}H\right) \left[ t;\rho ,w;\rho ^{\prime },w^{\prime
};\left\langle \omega \cdot \omega ^{\prime }\right\rangle \right] =
\end{equation*}
\vspace{-0.85cm} 
\begin{equation*}
i\sqrt{-t}\int_{0}^{\infty }\rho ^{\prime \prime d-1}d\rho ^{\prime \prime
}\int_{-\infty }^{\infty }dw^{\prime \prime }G\left[ t;\rho ^{\prime \prime
},w^{\prime \prime }\right] \int_{\mathbb{S}_{d-1}}d\omega ^{\prime \prime }B%
\left[ t;\rho ,w;\rho'' ,w'';  
\left\langle \omega \cdot \omega'' 
\right\rangle \right]
\end{equation*}
\vspace{-0.85cm} 
\begin{equation}
\times H\left[ t;\rho ^{\prime \prime },w^{\prime \prime };\rho ^{\prime
},w^{\prime };\left\langle \omega ^{\prime \prime }\cdot \omega ^{\prime
}\right\rangle \right]  \tag{3.16}
\end{equation}
or equivalently, by using the mass variables and introducing the convolution
product $\ast $ of $SO\left( d\right) $-invariant kernels $a$ and $b$ on the
sphere $\mathbb{S}_{d-1},$ namely  
\begin{equation}
\left( a\ast b\right) \left( \left\langle \omega \cdot \omega ^{\prime
}\right\rangle \right) =\int_{\mathbb{S}_{d-1}}d\omega ^{\prime \prime
}a\left( \left\langle \omega \cdot \omega ^{\prime \prime }\right\rangle
\right) b\left( \left\langle \omega ^{\prime \prime }\cdot \omega ^{\prime
}\right\rangle \right) ,  \tag{3.17}
\end{equation}
\begin{equation*}
\underline{H}_{\left( \zeta ,\zeta ^{\prime },t\right) }\left( \left\langle
\omega \cdot \omega ^{\prime }\right\rangle \right) =\underline{B}_{\left(
\zeta ,\zeta ^{\prime },t\right) }\left(\left\langle \omega \cdot \omega
^{\prime }\right\rangle \right) 
\end{equation*}
\vspace{-0.75cm} 
\begin{equation}
-i\int_{\Delta _{t}}
\left( \underline{B}%
_{\left( \zeta ,\zeta ^{\prime \prime },t\right) }\ast \underline{H}_{\left(
\zeta'' ,\zeta ^{\prime },t\right) }\right) \left( \left\langle \omega
\cdot \omega ^{\prime }\right\rangle \right)   
\underline{G}\left( \zeta ^{\prime \prime }\right) 
d\mu _{t}\left( \zeta ^{\prime \prime }\right).  
\tag{3.18}
\end{equation}

Let us now introduce the ``partial-wave expansion'' of invariant kernels 

\noindent $a\left( \left\langle \omega \cdot \omega ^{\prime }\right\rangle
\right) \equiv a\left( \cos \theta \right) $ on the sphere $\mathbb{S}_{d-1}$ by
the following formulae: 
\begin{equation}
a\left( \cos \theta \right) =\dfrac{1}{\omega _{d}}\sum\limits_{0\leq
\ell <\infty }\tilde{a}_{\ell }\  {\rm h}_d\left( \ell \right)  P_{\ell
}^{\left( d\right) }\left( \cos \theta \right) ,  \tag{3.19}
\end{equation}%
\begin{equation}
\tilde{a}_{\ell }=\omega _{d-1}\int_{-1}^{+1}P_{\ell }^{\left( d\right)
}\left( \cos \theta \right) a\left( \cos \theta \right) \left( \sin \theta
\right) ^{d-3}d\cos \theta  \tag{3.20}
\end{equation}%
where the functions $P_{\ell }^{\left( d\right) }$ are the ``ultraspherical
Legendre polynomials'', given by the following integral representation: 
\begin{equation}
P_{\ell }^{\left( d\right) }\left( \cos \theta \right) =\dfrac{\omega _{d-2}%
}{\omega _{d-1}}\int_{0}^{\pi }\left( \cos \theta +i\sin \theta \cos \varphi 
\right) ^{\ell }\left( \sin \varphi \right) ^{d-3}d\varphi ,  \tag{3.21}
\end{equation}%
\begin{equation}
{\rm h}_{d}\left( \lambda \right) =\dfrac{\left( 2\lambda +d-2\right) }{%
\left( d-2\right) !}\,\dfrac{\Gamma \left( \lambda +d-2\right) }{\Gamma
\left( \lambda +1\right) }  \tag{3.22}
\end{equation}%
and $\omega _{d}=2{\pi ^{d/2}\over\Gamma (d/2)}$  
is the area of the sphere $\mathbb{S}_{d-1}.$

Eq. (3.20) allows us to introduce the ``$t-$channel partial-waves'' \break 
$ \tilde h_{\ell}[t; \rho,w,\rho',w'] \equiv \tilde h_{\ell} (\zeta,\zeta',t)$ 
of the
restriction of the four-point function $H\left( \left[ k\right] \right) $
to $E_{d+1}^3:$%
\begin{equation}
\tilde{h}_{\ell }\left( \zeta ,\zeta ^{\prime },t\right) =\omega
_{d-1}\int_{-1}^{+1}P_{\ell }^{\left( d\right) }\left( \cos \theta
\right) \underline{H}_{\left( \zeta ,\zeta ^{\prime },t\right) }\left( \cos
\theta \right) \left( \sin \theta \right) ^{d-3}d\cos \theta , \tag{3.23}
\end{equation}
and similarly for  
$B\left( \left[ k\right] \right) $:
\begin{equation}
\tilde{b}_{\ell }\left( \zeta ,\zeta ^{\prime },t\right) =\omega
_{d-1}\int_{-1}^{+1} P_{\ell }^{\left( d\right) }\left(
\cos\theta \right) \underline{B}_{\left( \zeta ,\zeta ^{\prime
},t\right) }\left( \cos \theta \right) \left( \sin \theta \right)
^{d-3}d\cos \theta .  \tag{3.24}
\end{equation}

By now using the ``factorization property''\footnote{%
Note that the normalizations chosen for writing the definitions (3.17) and
(3.20) of the convolution-product and of the partial waves yield Eq. (3.25)
without extra-coefficient; they however differ by a factor $\omega _{d}$
from the standard normalization.} according to which the partial waves of $%
\left( {a\ast b}\right) \left( \left\langle \omega \cdot \omega
^{\prime }\right\rangle \right) $ are: 
\begin{equation}
( \widetilde{a\ast b}) _{\ell }=\widetilde{a}_{\ell }\widetilde{b}%
_{\ell },  \tag{3.25}
\end{equation}
we can replace the version (3.18) of the Bethe--Salpeter equation by the
following set of ``\textit{Bethe--Salpeter equations for the partial waves}%
'' of $H\left( \left[ k\right] \right) :$%
\begin{equation}
\tilde{h}_{\ell }\left( \zeta ,\zeta ^{\prime },t\right) =\tilde{b}%
_{\ell }\left( \zeta ,\zeta ^{\prime },t\right) +\int_{\Delta _{t}}
\tilde{b}_{\ell }\left( \zeta ,\zeta'' ;t\right) 
\tilde{h}_{\ell }\left( \zeta'' ,\zeta ^{\prime };t\right)  
\underline{G}\left( \zeta ^{\prime
\prime }\right) 
d\mu_{t}\left( \zeta ^{\prime \prime }\right).   
\tag{3.26}
\end{equation}

Note that each of these integral equations is comparable to a Bethe--Salpeter
equation for field theory in two-dimensional space-time. In view of the bounds
(3.9) for $H$ and (3.12) for $G$,  
the integral equations (3.15) and (3.26) appear as
Fredholm resolvent equations, to which the results of the Appendix apply   
(see Proposition A1 with $z,z'$ varying in $\Gamma_0$ for Eq (3.15) and the remark
after Proposition A3 for Eqs (3.26)).

\subsection{Bethe--Salpeter equation for absorptive parts in the mass
variables\ and\ angular\ variables: perikernel\newline
structure}

In section 2, we have recalled the fact that the axiomatic analyticity
domain of $H\left( \left[ k\right] \right) $ (or $B\left( \left[ k\right]
\right) )$ contains the Euclidean subspace of complex momentum-space and
provides a connection between the latter and the \textit{real}
Minkowskian subspace by travelling in the complex energy variables $%
K^{(0)},Z^{(0)},{Z'}^{(0)}$ at fixed $\vec{K},\vec{Z},\vec{Z}^{\prime }.$
This allowed one to reach the spectral sets $\Sigma_{s}, \Sigma_u,$  
to compute the corresponding discontinuities of $B \circ_{t} H$
and thereby to obtain the Bethe--Salpeter equations (2.6) and (2.7) for the
absorptive parts of $H\left( \left[ k\right] \right) $ in the $s$ and $u$%
-channels. This derivation was valid for arbitrary vectors $K=(K^{(0)}$ $%
\mathrm{complex,}$ $\vec{K}$ $\mathrm{real}).$

For $K=( 0,\vec{K}) ( K^{2}=t<0) ,$ an analyticity
property of similar type, but actually different since \textit{adapted to
the mass variables and angular variables} $\left( \zeta ,z\right) ,\left(
\zeta ^{\prime },z^{\prime }\right) ,$ was established in [1].In fact,
it was proven there that for all (real) values of $\left( \zeta ,\zeta
^{\prime }\right) $ in $\Delta _{t}\times \Delta _{t},$ the ``\textit{%
enlarged}'' axiomatic analyticity domain of $H\left( \left[ k\right] \right) 
,$ obtained by geometrical
techniques of analytic completion, contains the whole complex manifold $%
\hat{\Omega}_{(\zeta,\zeta' ,K)}$ parametrized by Eqs. (3.1)
...(3.6), in which $\left( z,z^{\prime }\right) $ varies on $X_{d-1}^{\left(
c\right) }\times X_{d-1}^{\left( c\right) },$ deprived from ``cuts'' generated
by the spectral sets $\Sigma_{s}\ ( ( Z-Z^{\prime })
^{2}\geq s_{0}) $ and $\Sigma_{u}\ ( ( Z+Z^{\prime
}) ^{2}\geq u_{0}) .$ In other words, for each $\left( \zeta
,\zeta ^{\prime }\right) $ fixed, the function $\underline{H}_{\left( \zeta
,\zeta ^{\prime },t\right) }\left( -z\cdot z^{\prime }\right) \equiv
H\left( \left[ k\right] \right) $ 
considered in subsection 3.1 as an invariant kernel
on the sphere 
$\mathbb{S}_{d-1}$ 
admits \textit{an analytic continuation on }%
$X_{d-1}^{\left( c\right) }\times X_{d-1}^{\left( c\right) }$ deprived from
cuts $\underline{\Sigma }_{s}\left( \zeta ,\zeta ^{\prime },t\right) $ and $%
\underline{\Sigma }_{u}\left( \zeta ,\zeta ^{\prime },t\right) $ (see Eqs.
(3.29), (3.30) below) which describe the traces of the spectral sets $%
\Sigma_{s}$ and $\Sigma_{u}$ in the manifold 
$\hat{\Omega}%
_{(\zeta ,\zeta ^{\prime },K)}.$ 
Moreover, since $\underline{H}_{_{\left(
\zeta ,\zeta ^{\prime },t\right) }}$ only depends on $z,z^{\prime }$ through
the variable $\cos \Theta = -z\cdot z^{\prime },$ it is analytic with respect to
this variable in the image of $X_{d-1}^{\left( c\right) }\times
X_{d-1}^{\left( c\right) }\backslash ( \underline{\Sigma }_{s}\left(
\zeta ,\zeta ^{\prime },t\right) \cup \underline{\Sigma }_{u}\left( \zeta
,\zeta ^{\prime },t\right) ) ,$ which is a \textit{cut-plane} of the
form $\mathbb{C}\backslash \left\{ \left[ \cosh v_{s}+\infty \left[\  \cup %
\ \right] -\infty ,-\cosh v_{u}\right] \right\} .$

The class of functions ${\cal K}(z,z')$ holomorphic in the previous cut-domains of 
$X_{d-1}^{\left( c\right) }\times X_{d-1}^{\left( c\right) }$ 
(also denoted by ${\cal K}(-z\cdot z')$ when they are Lorentz invariant) 
has been extensively studied in [11,12] under the name of 
\textit{(invariant) perikernels,} and their discontinuities have been
characterized as \textit{(invariant) }``\textit{Volterra kernels}'' on the
one-sheeted hyperboloid $X_{d-1};$ useful results involving these notions
will be recalled below.

Such a perikernel structure is satisfied not only by $H([k])$ but also by the
BS-kernel $B([k])$, since the latter enjoys the same analyticity domain as 
$H([k])$ for fixed $K$ (with $t=K^2 <0$); so we shall put similarly
$B([k]) \equiv {\underline B}_{(\zeta,\zeta',t)}(-z\cdot z').$ This perikernel 
structure of 
${\underline B}_{(\zeta,\zeta',t)}(-z\cdot z')$ will also be 
reobtained below in two ways by introducing and solving appropriate extensions of 
the BS-equations (3.18) and (3.26). 

In the continuation of this program, a basic role is played by the bounds (2.1) 
which $H([k])$ is assumed to satisfy in its axiomatic domain, in particular 
in the sets $\underline D_{(w,w',\rho')}$ of Propositions 3, 4 of [1]. In fact,  
it has been established in theorem 1 of [1] that bounds of the following form  
hold in each submanifold 
$\hat{\Omega}%
_{(\zeta ,\zeta ^{\prime },K)}:$ 
\begin{equation}
|{\underline H}_{(\zeta,\zeta',t)}(\cos \Theta_t)| \leq  
C^{(H)}_{(\zeta,\zeta',t)} \ {\rm e}^{N_H |\Im m \Theta_t|} 
|\sin \Re e \Theta_t|^{-n_H}. 
\tag{3.27}
\end{equation}
In the latter, $n_H$ describes the maximal local order of singularity of 
the distribution-boundary-values 
of ${\underline H}_{(\zeta,\zeta',t)}$
on the reals. We shall consider here $n_H$ as being independent 
\footnote{The occurrence of $\max (n_H, N_H)$ in place of $N_H$ in the exponential 
factor obtained in the bound (3.25) of [1] is in fact without physical content} 
of the exponent  
$N_H$ which governs the behaviour of $H$ at infinity according to 
the assumed bounds (2.1). 
Concerning the ``constant''
$C^{(H)}_{(\zeta,\zeta',t)},$ one can check (by following the proof of theorem 1 of [1]) 
that it can be taken equal to the uniform bound of $H$ in the Euclidean subspace, 
namely (see (3.9)):
\begin{equation}
C^{(H)}_{(\zeta,\zeta',t)} =   
C_E^{(H)} [(1+|t|^{1\over 2}) \ (1+\rho)\ (1+ |w||t|^{1\over 2})
 \ (1+\rho')\ (1+ |w'||t|^{1\over 2})]^{N_H}  . 
\tag{3.28}
\end{equation}

\vskip 0.3cm
\noindent
{\bf Absorptive parts: passage to the mass variables and hyperbolic angular variables; 
convolution products $\lozenge$}

\vskip 0.3cm

\noindent
The absorptive parts $\Delta _{s}H,\Delta _{u}H$ (resp. 
$\Delta _{s}B,\Delta _{u}B)$ defined in section\ 2 also appear at fixed $%
\left( \zeta ,\zeta ^{\prime },t\right) $ (or $\left( \rho ,w,\rho ^{\prime
},w^{\prime },t\right) )$ as being equal to the discontinuities 
$\Delta _{s}\underline{H}
_{( \zeta ,\zeta' ,t) }( -z\cdot z')$  
and $\Delta _{u}\underline{H}
_{( \zeta ,\zeta' ,t) }( -z\cdot z')$  
of $\underline{H}
_{( \zeta ,\zeta' ,t) }( -z\cdot z')$  
(and similarly for 
$\underline{B}
_{( \zeta ,\zeta' ,t) }$).  
The supports of these discontinuities, obtained by writing the
conditions $s=\left( Z-Z^{\prime }\right) ^{2}\geq s_{0},u=\left(
Z+Z^{\prime }\right) ^{2}\geq u_{0}$ in terms of the parametrization (3.1),
(3.2) are respectively:%
\begin{equation*}
\underline{\Sigma }_{s}\left( \zeta ,\zeta',t\right) =\  
\{(z,z')\in X_{d-1}^{(c)}\times X_{d-1}^{(c)};\   -z\cdot z' =\cosh v\geq \cosh v_{s}\},
\end{equation*}%
\begin{equation}
\mathrm{with\ }\cosh v_{s}=1+\frac{s_{0}+\left( \rho -\rho ^{\prime }\right)
^{2}-\left( w-w^{\prime }\right) ^{2}t}{2\rho \rho ^{\prime }}  \tag{3.29}
\end{equation}
\begin{equation*}
\mathrm{(or\ }\Theta _{t}=iv,\qquad v\geq v_{s});
\end{equation*}%
\vskip 0.3cm%
\begin{equation*}
\underline{\Sigma }_{u}\left( \zeta ,\zeta',t\right) =\  
\{(z,z')\in X_{d-1}^{(c)}\times X_{d-1}^{(c)};\   z\cdot z' =\cosh v\geq \cosh v_{u}\},
\end{equation*}%
\begin{equation}
\mathrm{with\ }\cosh v_{u}=1+\frac{u_{0}+\left( \rho -\rho ^{\prime }\right)
^{2}-\left( w+w^{\prime }\right) ^{2}t}{2\rho \rho ^{\prime }}  \tag{3.30}
\end{equation}%
\begin{equation*}
\mathrm{(or\ }\Theta _{t}=\pi +iv,\qquad v\geq v_{u}).
\end{equation*}

It is easy to check that these sets are respectively contained in the
regions $\left( z-z^{\prime }\right) ^{2}>0$ and $\left( z+z^{\prime
}\right) ^{2}>0.$ For $z,z^{\prime }$ real, i.e. belonging to the real
one-sheeted hyperboloid $X_{d-1},$ these relations express the fact that $z$
is either in the ``future'' or in the ``past'' of 
$z^{\prime }$ (resp.$-z^{\prime }).$ (Note that
the future and the  past of $z^{\prime },$ namely the regions $\Gamma ^{\pm
}\left( z^{\prime }\right) =X_{d-1}\cap \left\{ z\in \mathbb{R}%
^{d-1},z-z^{\prime }\in V^{\pm }\right\} $ are bounded by the cone of
generatrices of $X_{d-1}$ passing through $z^{\prime }).$ We conclude that
the trace of $\underline{\Sigma }_{s}\left( \zeta ,\zeta ^{\prime },t\right) $
on $X_{d-1}\times X_{d-1}$ is composed of two disjoint sets $\underline{\Sigma }%
_{s}^{+}\left( \zeta ,\zeta ^{\prime },t\right) $ and $\underline{\Sigma }%
_{s}^{-}\left( \zeta ,\zeta ^{\prime },t\right) ,$ corresponding
respectively to the conditions $z-z^{\prime }\in {\bar V}^{+}$ and $z-z^{\prime
}\in {\bar V}^-.$
Of course, these sets represent the supports of the respective components $%
\Delta _{s}^{+}H$ and $\Delta _{s}^{-}H$ of $\Delta _{s}H$ in the manifold $%
\hat{\Omega}_{\left( \zeta ,\zeta ^{\prime },K\right) },$ and this leads one
to distinguish two kernels $\Delta _{s}^{+}\underline{H}_{\left( \zeta
,\zeta ^{\prime },t\right) }\left( z,z^{\prime }\right) \equiv \Delta
_{s}^{+}H\left( \left[ k\right] \right) $ and $\Delta _{s}^{-}H_{\left(
\zeta ,\zeta ^{\prime },t\right) }\left( z,z^{\prime }\right) \equiv \Delta
_{s}^{-}H\left( \left[ k\right] \right) $ of disjoint supports on $X_{d-1},$
although (in view of Lorentz invariance) they are both represented \textit{%
by the same function of one variable} denoted earlier by $\Delta _{s}%
\underline{H}_{\left( \zeta ,\zeta ^{\prime },t\right) }\left( -z\cdot z^{\prime
}\right) .$ Kernels on $X_{d-1}$ such as $\Delta _{s}^{+}\underline{H}%
_{\left( \zeta ,\zeta ^{\prime },t\right) }\left( z,z^{\prime }\right) ,$
whose support is contained in the set $\left\{ \left( z,z^{\prime }\right)
\in X_{d-1}\times X_{d-1};z-z^{\prime }\in \bar{V}^{+}\right\} $ have been
introduced in [14] under the name of ``\textit{Volterra
kernels}'' on $X_{d-1}.$ One introduces similarly the kernels $\Delta _{u}^{\pm}%
\underline{H}_{\left( \zeta ,\zeta ^{\prime },t\right) }\left( z,z^{\prime }%
\right) $ with respective supports $\underline{\Sigma }_{u}^{\pm }\left( \zeta
,\zeta ^{\prime },t\right) $ distinguished by the
conditions $\pm \left( z+z^{\prime }\right)\in {\bar V}^{+},$ which are represented by
the same function $\Delta _{u}\underline{H}_{\left( \zeta ,\zeta ^{\prime
},t\right) }\left( -z\cdot z^{\prime }\right) .$

These considerations will now allow us to reinterpret the Bethe--Salpeter
equations for the absorptive parts (2.6) and (2.7), after rewriting the
latter in terms of the mass variables $(\zeta,\zeta' ) $ and of the ``hyperbolic  
angular variables'' $z,z^{\prime },$ varying on $X_{d-1}.$
As a counterpart of our expression (3.16) of $\left( B \circ_{t}H\right) $ in
Euclidean space, we can in fact rewrite the corresponding composition
product $\Delta _{s}B\,\Diamond _{t}\,\Delta _{s}H$ (see Eq. (2.10)) as
follows, in the situation where $Z$ is in the future of $Z^{\prime }$ (a
similar expression would be obtained in the opposite situation):%
\begin{equation*}
\left( \Delta _{s}B\,\Diamond _{t}\,\Delta _{s}H\right) \left[ t;\rho
,w;\rho ^{\prime },w^{\prime };-z\cdot z^{\prime }\right] =
\end{equation*}%
\begin{equation*}
\sqrt{-t}\int_{0}^{\infty }\rho ^{\prime \prime d-1}d\rho ^{\prime \prime
}\int_{-\infty }^{+\infty }G\left[ t;\rho ^{\prime \prime
},w^{\prime \prime }\right]dw'' \int\limits_{\Diamond \left( z, z^{\prime
}\right) \cap X_{d-1}}\Delta _{s} B\left[ t;\rho
,w;\rho^{\prime \prime },w^{\prime \prime };-z\cdot z'' 
\right] \times
\end{equation*}%
\begin{equation}
\cdots \Delta _{s} H\left[ t;\rho ^{\prime \prime },w^{\prime \prime };\rho
^{\prime },w^{\prime };-z^{\prime \prime }\cdot z^{\prime }\right] dz'', 
\tag{3.31}
\end{equation}%
where $dz^{\prime \prime }$ denotes the Lorentz-invariant measure $%
dz^{\prime \prime }=\dfrac{dz_{0}...dz_{d-2}}{z_{d-1}}$ on $X_{d-1}.$ The
fact that the integration region on $X_{d-1}$ is restricted to the
double-cone $\lozenge \left( z,z^{\prime }\right) $ is a consequence of the
integration prescription $\left\{ Z^{\prime \prime }\in \lozenge \left(
Z,Z^{\prime }\right) \right\} $ 
expressing the support properties of 
$\Delta_s B(K;Z,Z')$ and  
$\Delta_s H(K;Z,Z')$   
in Eq. (2.10). This results from the
implication relation $Z-Z^{\prime }\in \overline{V}^{+} \Longrightarrow 
z-z' \in {\bar V}^+,$ obvious from the
following identities 
(entailed by Eqs. (3.1), (3.2)):  
\begin{equation}
\rho \rho ^{\prime }\left( z-z^{\prime }\right) ^{2}=\left( Z-Z^{\prime
}\right) ^{2}+\left( \rho -\rho ^{\prime }\right) ^{2}+\left( w-w^{\prime
}\right) ^{2}|t|  \tag{3.32}
\end{equation}%
and%
\begin{equation}
\left( Z-Z^{\prime }\right) \cdot \left( z-z^{\prime }\right) =\left( \rho
+\rho ^{\prime }\right) \times \frac{\left( z-z^{\prime }\right) ^{2}}{2} 
\tag{3.33}
\end{equation}%
Moreover the same support properties of 
$\Delta_s B$ and  
$\Delta_s H$ also imply that the integrand at the r.h.s. of Eq.(3.31) vanishes 
outside a compact subset of the space of integration variables $(\rho'',w'')$; 
the integral (3.31) is therefore well-defined independently of the bounds on 
$\Delta_s B$ and  
$\Delta_s H$.  

It is now appropriate to introduce the notion of \textit{convolution product 
}$\lozenge$ of Volterra kernels on the one-sheeted hyperboloid (see 
[14,11]) by the following formula:
\begin{equation}
\left( F_{1}\Diamond F_{2}\right) \left( z,z^{\prime }\right)
=\int\limits_{\Diamond \left( z,z^{\prime }\right) \cap X_{d-1}}
F_{1}\left( z,z^{\prime \prime }\right) F_{2}\left( z^{\prime \prime
},z^{\prime }\right) dz'' .  \tag{3.34}
\end{equation}
It is to be noted that, due to the compactness of the integration region in 
(3.34), this convolution product remains meaningful for distribution-like 
Volterra kernels. 

For invariant Volterra kernels, which are of the form 
$F_{i}\left( z,z^{\prime }\right) =f_{i}\left( -z\cdot z^{\prime }\right)$, 
$i=1,2$ (as it is the case here for $\Delta _{s}^{+}\underline{H}_{\left(
\zeta ,\zeta ^{\prime },t\right) },\Delta _{s}^{+}\underline{B}_{\left(
\zeta ,\zeta ^{\prime },t\right) }$), formula (3.34) takes the following
alternative form:
\begin{equation}
\left( F_{1}\lozenge F_{2}\right) \left( z,z^{\prime }\right)  \equiv
\left( f_{1}\blacklozenge f_{2}\right) \left( \cosh v\right) = \frac{%
2\omega _{d-2}}{{\sinh v}^{d-3}}\int\limits_{\substack{ v_{1}\geq
0,v_{2}\geq 0 \\ v_{1}+v_{2}\leq v}}
f_{1}\left( \cosh v_{1}\right) f_{2}\left( \cosh v_{2}\right)  
\notag 
\end{equation}
\begin{equation}
\cdots \left[ \left( \cosh v-\cosh \left( v_{1}+v_{2}\right) \right) \left( \cosh
v-\cosh \left( v_{1}-v_{2}\right) \right) \right]^{d-4 \over 2} 
\ d\left( \cosh v_{1}\right) d\left( \cosh
v_{2}\right).   
\tag{3.35}
\end{equation}
Using the latter and putting   
$\Delta _{s}H\left[ t;\rho ,w;\rho ^{\prime },w^{\prime },-z\cdot z^{\prime
}\right] \equiv \Delta _{s}\underline{H}_{\left( \zeta ,\zeta ^{\prime
},t\right) }\left( -z\cdot z^{\prime }\right)$  
(and similarly for $\Delta _{s}B),$ we can then rewrite Eq. (3.31) as
follows:
\begin{equation*}
\left( \Delta _{s} B\ \lozenge _{t}\ \Delta _{s} H\right) \left[ t;\rho ,w;\rho
^{\prime },w^{\prime };-z\cdot z^{\prime }\right] =
\end{equation*}%
\begin{equation}
\int\limits_{\Delta _{t}}
\left( \Delta _{s}
\underline{B}_{\left( \zeta ,\zeta ^{\prime \prime },t\right) }\ \blacklozenge\   
\Delta _{s}\underline{H}_{\left( \zeta ^{\prime \prime },\zeta ^{\prime
},t\right) }\right) \left( -z\cdot z^{\prime }\right)   
\underline{G}\left( \zeta^{\prime \prime }\right) 
d\mu _{t}\left( \zeta^{\prime \prime }\right).   
\tag{3.36}
\end{equation}%
Similar expressions could be written for the three other composition
products $\Delta _{u}B\ \lozenge _{t}\ \Delta _{u}H,\ \Delta _{s}B\ \lozenge 
_{t}\  \Delta_u H,\ \Delta _{u}B\ \lozenge _{t}\ \Delta _{s}H$ 
of Eqs. (2.6), (2.7), in terms
of corresponding convolution products of Volterra kernels.  

\vskip 0.3cm 
\noindent
{\bf From the Euclidean BS-equation to the BS-equation for 
$s$ and $u-$channel absorptive parts 
through contour-distortion of ``perikernel convolution products''} 

\vskip 0.3cm 
We now recall the basic relationship 
which relates the $\ast $-convolution product of kernels on the
``Euclidean'' sphere $\mathbb{S}_{d-1}$  and the $\lozenge-$convolution product 
of Volterra kernels on the one-sheeted hyperboloid $X_{d-1}$
(theorem 2' of [11]). 

\vskip 0.3cm 
{\sl Being given two perikernels $\mathcal{K}_{i}\left( z,z^{\prime
}\right) ,i=1,2,$ on $X_{d-1}^{\left( c\right) }$ whose respective
discontinuities on the sets $\left\{ z-z^{\prime }\in V^{\pm }\right\}
,\left\{ z+z^{\prime }\in V^{\pm }\right\} $ are the Volterra kernels $%
\Delta _{s}^{\pm }\mathcal{K}_{i}\left( z,z^{\prime }\right) ,\Delta
_{u}^{\pm }\mathcal{K}_{i}\left( z,z^{\prime }\right) ,$ there exists a
perikernel $\mathcal{K}$ denoted by $\mathcal{K}=\mathcal{K}_{1}\ast
^{\left( c\right) }\mathcal{K}_{2},$ such that:

i) the restrictions of $\mathcal{K},\mathcal{K}_{1},\mathcal{K}_{2},$
to the ``Euclidean'' sphere $\mathbb{S}_{d-1}$ are such that:
\begin{equation}
\mathcal{K}_{|\mathbb{S}_{d-1}}={\mathcal{K}_1}_{|{\mathbb{S}_{d-1}}}{\ast
}\ {\mathcal{K}_2}_{|\mathbb{S}_{d-1}},  \tag{3.37}
\end{equation}

ii) the discontinuities $\Delta _{s}^{+}\mathcal{K}\left( z,z^{\prime
}\right) $ and $\Delta _{u}^{+}\mathcal{K}\left( z,z^{\prime }\right) $ are
given by the following $\lozenge-$convolution products:
\begin{equation*}
\Delta _{s}^{+}\mathcal{K}=\Delta _{s}^{+}\mathcal{K}_{1}\lozenge \Delta
_{s}^{+}\mathcal{K}_{2}+\Delta _{u}^{-}\mathcal{K}_{1}\lozenge \Delta
_{u}^{+}\mathcal{K}_{2}
\end{equation*}%
\begin{equation}
\Delta _{u}^{+}\mathcal{K}=\Delta _{s}^{-}\mathcal{K}_{1}\lozenge \Delta
_{u}^{+}\mathcal{K}_{2}+\Delta _{u}^{+}\mathcal{K}_{1}\lozenge \Delta
_{s}^{+}\mathcal{K}_{2},  \tag{3.38}
\end{equation}%
(similar formulae being satisfied by $\Delta _{s}^{-}\mathcal{K}$ and $%
\Delta _{u}^{-}\mathcal{K}$), 

iii) for every $(z,z')$ in the analyticity domain of 
$\mathcal{K},$ there exists a class of cycles $\Gamma (z,z')$ such that 
\begin{equation}
\mathcal{K}(z,z') =\int_{
\Gamma(z,z')}   
\mathcal{K}_{1}(z,z'') 
\mathcal{K}_{2}(z'',z') dz'',  
\tag{3.39}
\end{equation}
$\Gamma(z,z')$ being 
obtained by continuous distortion inside the analyticity domain of 
the integrand from the special cycle  
$\Gamma_0(z,z')\equiv  
\mathbb{S}_{d-1}$, relevant for the  Euclidean configurations 
$(z,z') \in
\mathbb{S}_{d-1} 
\times \mathbb{S}_{d-1},$  

iv) if  
$\mathcal{K}_{1},\mathcal{K}_{2}$ are invariant perikernels,  
$\mathcal{K}$ is also an invariant perikernel.} 

\vskip 0.3cm 
This statement can then be applied for each set of fixed values of $\left(
\zeta ,\zeta ^{\prime },\zeta'', t\right)$ to the (invariant) perikernels 
$ \underline{B}_{( \zeta ,\zeta'',t)}  
( -z\cdot z'')  $ and $\underline{H}_{\left(
\zeta'' ,\zeta' ,t\right) }( -z''\cdot z')  .$ 
It implies the existence of an (invariant)
perikernel $\left( \underline{B}_{\left( \zeta ,\zeta'' ,t\right)
}\ast ^{\left( c\right) }\underline{H}_{\left( \zeta'' ,\zeta' 
,t\right) }\right) (-z\cdot z') ,$ whose restriction to the
``Euclidean'' sphere $\mathbb{S}_{d-1}$ is correspondingly the kernel 
$\left( {\underline B}_{(\zeta,\zeta'',t)}  
\ast {\underline H}_{( \zeta'' ,\zeta' ,t) }\right) $
appearing at the r.h.s. of Eq. (3.18).  

In view of property iii), one can see that {\sl the Bethe--Salpeter equation (3.18) can be
analytically continued} to all $\left( z,z^{\prime }\right) $ in $%
X_{d-1}^{\left( c\right) }\times X_{d-1}^{\left( c\right) },$ for all $%
t<0,\left( \zeta ,\zeta ^{\prime }\right) $ in $\Delta _{t}\times \Delta _{t}
,$ under the following form:
\begin{equation}
\underline{H}_{\left( \zeta ,\zeta ^{\prime },t\right) }\left( z,z^{\prime }%
\right) =\underline{B}_{\left( \zeta ,\zeta ^{\prime },t\right) }\left(  
z,z^{\prime }\right) -i\int\limits_{\Delta _{t}}
\left( \underline{B}_{\left( \zeta ,\zeta'',  
t\right) }\ast ^{\left( c\right) }\underline{H}_{\left( \zeta'' ,\zeta',  
t\right) }\right) \left( z,z^{\prime }\right)  
\underline{G}\left(
\zeta^{\prime \prime }\right) 
d\mu _{t}(\zeta'').
\tag{3.40}
\end{equation}
The latter can be considered as a Fredholm resolvent equation 
in complex space whose integration space is  
the product of $\Delta_t$
by the ``floating cycle'' $\Gamma$ on which the complex points 
$z,z'$ vary. In view of general results of [13] adapted to the present situation
in the Appendix, {\sl the function ${\underline B}_{(\zeta,\zeta',t)}(z,z')  $ 
is directly obtained with its full perikernel structure as the solution of the 
Fredholm equation (3.40)}: in fact, 
${\underline B}_{(\zeta,\zeta',t)}(z,z')  $ 
can be identified with the resolvent 
$R_{H|\alpha =-1}(t; \rho,w,\rho',w',z,z')$ of Propositions A1 and A2 ($F$ being replaced
by $H$). 
In the case when the boundary values of  
$\underline{H}_{\left( \zeta ,\zeta ^{\prime },t\right) }( z,z')$ 
and its discontinuities are continuous (namely if the bounds (3.27) hold 
with $n_H =0$), the Fredholm solution  
${\underline B}_{(\zeta,\zeta',t)}(z,z')  $ of Eq. (3.40) satisfies the same 
regularity properties and therefore one can apply the property ii) 
(Eq. (3.38)) of the $\ast^{(c)}-$composition product of perikernels
for computing side-by-side the discontinuities of Eq. (3.40). This yields: 
\begin{equation}
\Delta _{s}^{+}\underline{H}_{\left( \zeta ,\zeta ^{\prime },t\right)
}=\Delta _{s}^{+}\underline{B}_{\left( \zeta ,\zeta ^{\prime },t\right) }+ 
\tag{3.41}
\end{equation}%
\begin{equation*}
\int_{\Delta_t}  \left[ \Delta _{s}^{+}\underline{B}_{\left(
\zeta ,\zeta'' ,t\right) }\lozenge \Delta _{s}^{+}\underline{H}%
_{\left( \zeta'' ,\zeta' ,t\right) }+\Delta _{u}^{-}\underline{B}%
_{\left( \zeta ,\zeta'' ,t\right) }\lozenge \Delta _{u}^{+}%
\underline{H}_{\left( \zeta'' ,\zeta' ,t\right) }\right] 
\underline{G}\left(
\zeta ^{\prime \prime }\right) 
d\mu _{t}\left( \zeta ^{\prime \prime }\right) 
\end{equation*}%
\begin{equation} 
\Delta _{u}^{+}\underline{H}_{\left( \zeta ,\zeta ^{\prime },t\right)
}=\Delta _{u}^{+}\underline{B}_{\left( \zeta ,\zeta ^{\prime },t\right) }+
\tag{3.42}
\end{equation}
\begin{equation*}
\int_{\Delta_t} \left[ \Delta _{s}^{-}\underline{B}_{\left(
\zeta ,\zeta'' ,t\right) }\lozenge \Delta _{u}^{+}\underline{H}%
_{\left( \zeta'' ,\zeta' ,t\right) }+\Delta _{u}^{+}\underline{B}%
_{\left( \zeta ,\zeta'' ,t\right) }\lozenge \Delta _{s}^{+}%
\underline{H}_{\left( \zeta'' ,\zeta' ,t\right) }\right] 
\underline{G}\left(
\zeta ^{\prime \prime }\right) 
d\mu _{t}\left( \zeta ^{\prime \prime }\right) 
\end{equation*}
(with similar expressions for $\Delta _{s,u}^{-}\underline{H}_{\left( \zeta
,\zeta ^{\prime },t\right) }(z,z')$).   
Then by comparing the convolution products here obtained with the forms (3.31),(3.36)
of the $\lozenge_t-$composition-product (2.10),  
we see that the latter equations are in fact identical to Eqs (2.6),(2.7).  
So in this new presentation making use of perikernel-convolution-products,  
we have reobtained the BS-equations for the absorptive parts in terms of mass variables 
and hyperbolic angular variables. 

\vskip 0.2cm 
We now make use of the fact that both Volterra
kernels $\left( \Delta _{s}^{+}\underline{H}_{\left( \bullet \right)} 
,\Delta _{s}^{-}\underline{H}_{\left( \bullet \right) }\right) $ are 
represented by the same Lorentz-invariant function $\Delta _{s}\underline{H}%
_{\left( \bullet \right) }\left( -z\cdot z^{\prime }\right) ,$ with $%
-z\cdot z^{\prime }=\cosh v\geq 1,$ while both Volterra kernels $%
\left( \Delta _{u}^{+}\underline{H}_{\left( \bullet \right) },\Delta _{u}^{-}%
\underline{H}_{\left( \bullet \right) }\right) $ are represented by the same
function $\Delta _{u}\underline{H}_{\left( \bullet \right) }\left( -z\cdot   
z^{\prime }\right) ,$ with $-z\cdot z^{\prime }=-\cosh v\leq -1.$ It is then
convenient to put $\widehat{\Delta _{u}\underline{H}}_{\left( \bullet \right)
}\left( \cosh v\right) =\Delta _{u}\underline{H}_{\left( \bullet \right)
}\left( -\cosh v\right) ;$ similar considerations are done for $\underline{B}%
_{\left( \bullet \right) }.$ Then, 
Eqs. (3.41), (3.42) can be rewritten in terms
of $\blacklozenge-$convolution products of invariant Volterra kernels (see Eq.(3.35)) 
as follows:
\begin{equation}
\Delta _{s}\underline{H}_{\left( \zeta ,\zeta' ,t\right)} =  
\Delta _{s}\underline{B}_{\left( \zeta ,\zeta' ,t\right)} + \cdots  
\tag{3.43}
\end{equation}
\begin{equation}
\Delta _{s}\underline{B}_{\left( \zeta ,\zeta ^{\prime \prime },t\right)
}\blacklozenge \Delta _{s}\underline{H}_{\left( \zeta ^{\prime \prime },\zeta
^{\prime },t\right) }+\widehat{\Delta _{u}\underline{B}}_{\left( \zeta
,\zeta ^{\prime \prime },t\right) }\blacklozenge \widehat{\Delta _{u}\underline{H}}%
_{\left( \zeta ^{\prime \prime },\zeta ^{\prime },t\right) }  \notag
\end{equation}
and 
\begin{equation}
\widehat{\Delta _{u}\underline{H}}_{\left( \zeta ,\zeta' ,t\right)} =  
\widehat{\Delta _{u}\underline{B}}_{\left( \zeta ,\zeta' ,t\right)} + \cdots  
\tag{3.44}
\end{equation}
\begin{equation}
\Delta _{s}\underline{B}_{\left( \zeta ,\zeta ^{\prime \prime },t\right)
}\blacklozenge \widehat{\Delta _{u}\underline{H}}_{\left( \zeta'' ,\zeta' 
,t\right) }+\widehat{\Delta _{u}\underline{B}}_{\left( \zeta ,\zeta'',   
t\right) }\blacklozenge \Delta _{s}\underline{H}_{\left( \zeta
^{\prime \prime },\zeta ^{\prime },t\right) .}  \notag
\end{equation}
In these equations , all the terms are functions of the variable $\cosh v$ 
varying on the half-line $[1,+\infty [$. 

\vskip 0.3cm
\noindent
{\bf Symmetrized and antisymmetrized Bethe--Salpeter equations for the
absorptive parts:}

Let us put 
\begin{equation}
\Delta ^{\left( s\right)
}\underline{B}_{\left( \bullet \right) } =  
\Delta _{s}\underline{B}_{\left( \bullet \right) }+\widehat{%
\Delta _{u}\underline{B}}_{\left( \bullet \right) },  \ \  
\Delta ^{\left( s\right)
}\underline{H}_{\left( \bullet \right) } =  
\Delta _{s}\underline{H}_{\left( \bullet \right) }+\widehat{%
\Delta _{u}\underline{H}}_{\left( \bullet \right) },  \ \  
\tag{3.45}
\end{equation}%
and%
\begin{equation}
\Delta ^{\left( a\right)
}\underline{B}_{\left( \bullet \right) } =  
\Delta _{s}\underline{B}_{\left( \bullet \right) }-\widehat{%
\Delta _{u}\underline{B}}_{\left( \bullet \right) },  \ \  
\Delta ^{\left( a\right)
}\underline{H}_{\left( \bullet \right) } =  
\Delta _{s}\underline{H}_{\left( \bullet \right) }-\widehat{%
\Delta _{u}\underline{H}}_{\left( \bullet \right) },  \ \  
\tag{3.46}
\end{equation}
Then by adding-up and subtracting Eqs. (3.43) and (3.44) side by side 
one obtains the
following \textit{decoupled Bethe--Salpeter equations for the symmetric and
antisymmetric combinations of the $s$ and $u$-channel absorptive parts}:
\begin{equation}
\Delta ^{\left( s\right) }\underline{H}_{\left( \zeta ,\zeta ^{\prime
},t\right) }=\Delta ^{\left( s\right) }\underline{B}_{\left( \zeta ,\zeta
^{\prime },t\right) }+\int\limits_{\Delta _{t}}
\Delta ^{\left( s\right) }\underline{B}_{\left(
\zeta ,\zeta'',t\right) }\blacklozenge \Delta ^{\left( s\right) }\underline{H%
}_{\left( \zeta ^{\prime \prime },\zeta ^{\prime },t\right) }  
{\underline G}(\zeta'')
d\mu _{t}\left( \zeta
^{\prime \prime }\right) 
\tag{3.47}
\end{equation}%
and%
\begin{equation}
\Delta ^{\left( a\right) }\underline{H}_{\left( \zeta ,\zeta ^{\prime
},t\right) }=\Delta ^{\left( a\right) }\underline{B}_{\left( \zeta ,\zeta
^{\prime },t\right) }+\int\limits_{\Delta _{t}}
\Delta ^{\left( a\right) }\underline{B}_{\left( \zeta ,\zeta'',   
t\right) }\blacklozenge \Delta ^{\left( a\right) }\underline{H}_{\left( \zeta
^{\prime \prime },\zeta ^{\prime },t\right) }  
{\underline G}(\zeta'')
d\mu _{t}\left( \zeta
^{\prime \prime }\right).  
\tag{3.48}
\end{equation}
{\bf The case of distribution-like boundary values}  

\vskip 0.3cm
This more general case is characterized by a {\sl positive} exponent 
$n_H$ in the bound (3.27) on 
$\underline{H}_{( \zeta ,\zeta',t) }(\cos \Theta_t)$. Then 
the function 
$\underline{B}_{( \zeta ,\zeta',t) }(\cos \Theta_t)$ 
obtained as the Fredholm resolvent of 
$\underline{H}_{( \zeta ,\zeta',t) }$ from Eq. (3.40) cannot be 
proved directly to enjoy a similar power bound near the reals
(in view of the occurrence of the factor 
$|\sin \Re e \Theta_t|^{-n_H}$  in the constant $M_{\Gamma}$ inside 
the argument of the entire function $\Phi'$ in the bound (A.20)). 
The fact that such a power and the corresponding distribution character of the 
boundary values of 
$\underline{B}_{( \zeta ,\zeta',t) }$ still hold true (near the $s$ and $u-$cuts) 
is a consequence of the following properties:  

\vskip 0.2cm
a)\ For any Feynman convolution $F_1 \circ_t F_2$  and for any composition product 
of perikernels ${\cal K}_1 \ast^{(c)} {\cal K}_2$, the discontinuity formulae 
(2.12),(2.13) and (3.38) can still be justified (by the corresponding 
contour-distortion arguments) when the boundary values are not continuous but  
governed by power bounds near the reals. 
These formulae then hold as well-defined convolution-type products of 
distributions with supports in a salient cone (resulting in the double-cone-shaped 
integration region of the $\lozenge_t$ and $\lozenge-$products).   

\vskip 0.2cm
b)\ In view of the additivity property of spectral regions (see Sec 2), any 
four-point function of the form 
$H^{\circ_t (n+1)}= H\circ_t \cdots \circ_t H$ ($n$ products) has absorptive parts 
$\Delta_s H^{\circ_t (n+1)}$ and 
$\Delta_u H^{\circ_t (n+1)}$ admitting thresholds $s_n$ and $u_n$ ``of order $n$'':
to make it simple, in the case when $s_0 =u_0$ one has $s_n=u_n = (n+1)^2 s_0$.   

\vskip 0.2cm
c)\ For every $n$ let  
$B_{n+1}(K,Z,Z') $    
be the solution of the auxiliary equation 
$H^{\circ_t (n+1)} =    
B_{n+1}  +  (-1)^n  
B_{n+1} \circ_t     
H^{\circ_t (n+1)} $.    
$B_{n+1} $ has the same analyticity domain as    
$H^{\circ_t (n+1)} $, namely it has the same thresholds $s=s_n$, $u=u_n$. 
This results either from the 
Fredholm-type analysis of [8,9] or from 
the present one in the perikernel framework by using contours $\Gamma$ 
in $X_{d-1}^{(c)}$ (see [11] and the Appendix).  

\vskip 0.2cm
d)\ For every $n$ the following relation is obtained by iterating the 
equation $B= H- B\circ_t H$ and taking into account the defining equation of   
$B_{n+1}$:      
\begin{equation}
B= \sum_{p=1}^{n+1} (-1)^{p-1} 
H^{\circ_t p} +    
\sum_{p=1}^{n+1} (-1)^{n+p} 
B_{n+1} \circ_t H^{\circ_t p}. 
\tag{3.49}
\end{equation}

\vskip 0.2cm
Being interested in the boundary values and absorptive parts 
$\Delta_s B,\Delta_u B$ of $B$ in the real regions  
${\cal R}_s^{(n)}$, 
${\cal R}_u^{(n)}$ 
where (respectively)
$s= (Z-Z')^2 <s_n$ and 
$u= (Z+Z')^2 <u_n$, one can see that only the terms of the   
first sum at the r.h.s. of Eq. (3.49) will contribute to these 
absorptive parts. In fact in view of c) all the terms of the second 
sum in Eq. (3.49) are analytic in the regions  
${\cal R}_s^{(n)}$ 
and ${\cal R}_u^{(n)}$.  
Therefore in view of a) the existence of power bounds near the reals for $B$ 
and the formulae for the corresponding absorptive parts
in ${\cal R}_s^{(n)}$ and  
${\cal R}_u^{(n)}$ are completely governed by the finite sum  
$\sum_{p=1}^{n+1} (-1)^{p-1} 
H^{\circ_t p}$. It then also follows that one can apply a) 
directly to the Feynman convolution $B\circ_t H$ in these regions 
and therefore derive Eqs (3.41), (3.42) by applying the 
contour-distortion argument to Eq. (3.40) (as in the case of 
continuous boundary values).  

\subsection{Complex angular momentum diagonalization of the Bethe--Salpeter
equation}

We shall now apply the main result of [1] (theorem 5) to
the four-point function $H\left( \left[ k\right] \right) $.
This result relies basically on the Laplace-type $L_d-$transformation 
(see [12] for a complete study) which associates with each invariant 
Volterra kernel with moderate growth $F(z,z') \equiv f(-z\cdot z')$ on 
$X_{d-1}$ the following analytic function (see proposition III-3 of [12]):
\begin{equation}
{\tilde F}(\lambda) = \omega_{d-1} \int_0^{\infty} f(\cosh v) 
Q^{(d)}_{\lambda}(\cosh v) \ (\sinh v)^{d-2} dv.  
\tag{3.50}
\end{equation}
In this equation, 
$Q^{(d)}_{\lambda}(\cosh v)$ 
denotes the second-kind Legendre function in dimension $d$ whose integral 
representation is given by Eq.(4.36) of [1].  
If $|f(\cosh v)|$ is majorized by  ${\rm cst}\  {\rm e}^{Nv}$, 
then $\tilde F(\lambda)$ is proved to be 
holomorphic in the half-plane  
$\mathbb{C}_{+}^{(N)}= \{\lambda \in \mathbb{C};\  \Re e \lambda > N\}.$

According to theorem 5 of [1], the absorptive parts $\Delta
_{s}\underline H_{(\zeta,\zeta',t)}$ and 
$\Delta _{u}\underline H_{(\zeta,\zeta',t)}$ of a function
$\underline H_{(\zeta,\zeta',t)}(\cos\Theta_t)$ 
satisfying the bounds (3.27) admit ``$t$-channel Laplace-type
transforms'' ${\tilde H}_{s}\left( \zeta ,\zeta ^{\prime };t,\lambda
_{t}\right) $ and ${\tilde H}_{u}\left( \zeta ,\zeta ^{\prime };t,\lambda
_{t}\right) ,$ which are holomorphic with respect to the {\sl complex angular momentum 
variable} $\lambda _{t}$ in the half-plane $\mathbb{C}_{+}^{( N_H)}.$ 
These Laplace-type transforms, obtained by applying the $L_d-$transformation 
(3.50) at fixed values of $\zeta,\zeta',t$ are:  
\begin{equation}
{\tilde H}_{s}\left( \zeta ,\zeta ^{\prime };t,\lambda _{t}\right)
=\omega _{d-1}\int_{0}^{\infty }\Delta _{s}\underline{H}_{\left( \zeta
,\zeta ^{\prime },t\right) }\left( \cosh v\right) Q_{\lambda _{t}}^{\left(
d\right) }\left( \cosh v\right) \left( \sinh v\right) ^{d-2}dv  \tag{3.51}
\end{equation}
\begin{equation}
{\tilde H}_{u}\left( \zeta ,\zeta ^{\prime };t,\lambda _{t}\right)
=\omega _{d-1}\int_{0}^{\infty }\Delta _{u}\underline{H}_{\left( \zeta
,\zeta ^{\prime },t\right) }\left( -\cosh v\right) Q_{\lambda _{t}}^{\left(
d\right) }\left( \cosh v\right) \left( \sinh v\right) ^{d-2}dv  \tag{3.52}
\end{equation}
In the general case, these formulae have to be understood 
in the sense of distributions, 
namely with 
$Q_{\lambda _{t}}^{\left(
d\right) }\left( \cosh v\right) \left( \sinh v\right) ^{d-2} $ 
playing the role of a test-function. 
\vskip 0.2cm

Moreover, the  ``symmetric and antisymmetric Laplace-type transforms''
\begin{equation}
\tilde{H}^{\left( s\right) }=\tilde{H}_{s}+\tilde{H}_{u}\quad ,\quad \tilde{H%
}^{\left( a\right) }=\tilde{H}_{s}-\tilde{H}_{u},  \tag{3.53}
\end{equation}
which are defined in terms of $\Delta ^{\left( s\right) }\underline{H}_{\left( \zeta
,\zeta ^{\prime },t\right) }$ and $\Delta ^{\left( a\right) }\underline{H}%
_{\left( \zeta ,\zeta ^{\prime },t\right) }$ (see Eqs (3.45), (3.46)) via the
similar formulae:
\begin{equation}
\tilde{H}^{\left( s\right) ,\left( a\right) }\left( \zeta ,\zeta ^{\prime
};t,\lambda _{t}\right) =\omega _{d-1}\int_{0}^{\infty }
\Delta ^{\left(
s\right) ,\left( a\right) }\underline{H}_{\left( \zeta ,\zeta ^{\prime
},t\right) }\left( \cosh v\right) 
Q_{\lambda _{t}}^{\left( d\right) }\left(
\cosh v\right) \left( \sinh v\right) ^{d-2}dv  \tag{3.54}
\end{equation}
enjoy the following ``{\sl Froissart--Gribov-type equalities}''  
\begin{equation}
{\rm for} \ \ 2\ell >N_H \quad \quad 
\tilde{H}^{\left( s\right) }\left( \zeta ,\zeta ^{\prime
},t,2\ell \right) =
\tilde{h}_{2\ell }\left( \zeta ,\zeta ^{\prime
},t\right) 
\tag{3.55}
\end{equation}
\begin{equation}
{\rm for}\ \ 2\ell +1>N_H,\quad \quad 
\tilde{H}^{\left( a\right) }\left( \zeta ,\zeta ^{\prime
},t,2\ell +1 \right) =
\tilde{h}_{2\ell +1 }\left( \zeta ,\zeta ^{\prime
},t\right) 
\tag{3.56}
\end{equation}
$\tilde{H}^{\left( s\right) }$ and $\tilde{H}^{\left( a\right) }$
are {\sl Carlsonian} [5] (i.e. unique) {\sl interpolations} of the
respective sets of partial waves $\left\{ \tilde{h}_{2\ell };\ 2\ell
>N_H\right\} $ and $\left\{ \tilde{h}_{2\ell +1};\ 2\ell +1>N_H\right\} ;$ they
indeed satisfy bounds of the following form in $\mathbb{C}_{+}^{\left(
N_H+\varepsilon \right) }$ (for all positive $\varepsilon ,\varepsilon
^{\prime }):$%
\begin{equation}
\left| \tilde{H}^{\left( s\right) ,\left( a\right) }\left( \zeta ,\zeta
^{\prime };t,\lambda _{t}\right) \right| \leq 
c^{(s),(a)}_{(\varepsilon,\varepsilon')}\ C^{(H)}_{(\zeta,\zeta',t)} \  
\left| \lambda _{t}-N_H\right|
^{n_H-\frac{d-2}{2}+\varepsilon'}\  {\rm e}^ {-\left[ \Re e \lambda _{t}-\left(
N_H+\varepsilon \right) \right] v_{0}}  \tag{3.57}
\end{equation}%
where $v_{0}=\min \left( v_{s}\left( \zeta ,\zeta ^{\prime },t\right)
,v_{u}\left( \zeta ,\zeta ^{\prime },t\right) \right) ,v_{s},v_{u}$ being
the quantities introduced in (3.29), (3.30). 

\vskip 0.3cm
These bounds on $\tilde H^{(s)}$ and 
$\tilde H^{(a)}$ reexpress exactly (after symmetrization and antisymetrization) 
the bounds (4.84) established in theorem 5 of [1];
in particular, the occurrence of the power of $|\lambda_t - N_H|$ is correlated to 
the distribution character of 
$\Delta ^{\left(
s\right) ,\left( a\right) }\underline{H}_{\left( \zeta ,\zeta ^{\prime
},t\right) }\left( \cosh v\right)$ encoded in the bound (3.27).  
However, we now have 
rewritten the constants $C^{\varepsilon,\varepsilon'}_{s,u}$ of the latter reference 
under the form 
$c^{(s),(a)}_{(\varepsilon,\varepsilon')}\ C^{(H)}_{(\zeta,\zeta',t)} ,$ 
where the quantity 
$ C^{(H)}_{(\zeta,\zeta',t)} $ (see Eq (3.28)) contains 
the full dependence of the bound 
(3.27) with respect to the mass-variables $\zeta,\zeta'$, while   
$c^{(s)}_{(\varepsilon,\varepsilon')}$ and  
$c^{(a)}_{(\varepsilon,\varepsilon')}$ are 
purely numerical constants: the dependence of  
the bound (3.57) with respect to  
$ C^{(H)}_{(\zeta,\zeta',t)}, $ 
results from the linearity of the transformations which associate 
${\tilde H}^{(s)}(\zeta,\zeta';t,\lambda_t)$ and  
${\tilde H}^{(a)}(\zeta,\zeta';t,\lambda_t)$   
with ${\underline H}_{(\zeta,\zeta',t)}(\cos \Theta_t).$   

\vskip 0.2cm
Since 
${\tilde H}^{(s)}$ and  
${\tilde H}^{(a)}$ 
provide interpolations in the variable $\lambda_t$ of the respective sequences of 
{\sl even} and {\sl odd} Euclidean partial waves 
$\{\tilde h_{2\ell}(\zeta,\zeta',t);\ 2\ell > N_H\}$ and 
$\{\tilde h_{2\ell+1}(\zeta,\zeta',t);\ 2\ell +1 >N_H\}$  
of $H([k]),$ it is now natural to consider the corresponding interpolations in $\lambda_t$ 
(for $\Re e \lambda_t >N_H$) of the BS-equations (3.26) for these partial waves, which  
can be written as follows:  
\begin{equation*}
\tilde{H}^{\left( s\right) ,\left( a\right) }\left( \zeta ,\zeta ^{\prime
};t,\lambda _{t}\right) =\tilde{B}^{\left( s\right) ,\left( a\right) }\left(
\zeta ,\zeta ^{\prime };t,\lambda _{t}\right) +
\end{equation*}%
\begin{equation}
\int\limits_{\Delta _{t}}
\tilde{B}^{\left(
s\right) ,\left( a\right) }\left( \zeta ,\zeta'' ;t,\lambda
_{t}\right)  \tilde{H}^{\left( s\right) ,\left( a\right) }\left(
\zeta ^{\prime \prime },\zeta ^{\prime };t,\lambda _{t}\right) 
\underline{G}\left( \zeta ^{\prime \prime }\right) 
d\mu _{t}\left( \zeta ^{\prime \prime }\right).   
\tag{3.58}
\end{equation}
We have thus obtained 
two decoupled Bethe--Salpeter-type equations for  
$\tilde{B}^{\left( s\right) }$ and for $\tilde{B}^{\left( a\right) }$ in
terms of $\tilde{H}^{\left( s\right) }$ and $\tilde{H}^{\left( a\right) }$
respectively, in which the Fredholm integration space reduces to the
two-dimensional real region $\Delta _{t}$ of the plane of squared-mass
variables $\zeta ^{\prime \prime }=\left( \zeta _{1}^{\prime \prime },\zeta
_{2}^{\prime \prime }\right) ,$ while $\left( t,\lambda _{t}\right) $ are
parameters varying in 
$\mathbb{R}^{-}\times \mathbb{C}_{+}^{(N_H)}.$

In view of the mass-dependence  
of the uniform bounds (3.57) on  
${\tilde H}^{(s)}$ and  
${\tilde H}^{(a)},$ 
the Fredholm method can again be applied 
\footnote{Note that in the bound (3.57) the mass dependence has to be 
majorized by 
$ C^{(H)}_{(\zeta,\zeta',t)} $
since the exponential factor (also mass-dependent through $v_0$) has no 
better uniform majorant than $1$.  
In the treatment of the Appendix one then uses a majorant of 
$ C^{(H)}_{(\zeta,\zeta',t)} $
of the form (A.1) as explained there.}
in the equivalent, 
but simpler form specified in the Appendix,
where the {\sl variable} integration space $\{\zeta \in \Delta_t \}$ is replaced by the {\sl fixed}
integration space $\{(\rho'',w'')\in {\mathbb R}^+ \times {\mathbb R}\}$.  
This proves the existence of the solutions 
${\tilde B}^{(s)}(\zeta,\zeta';t,\lambda_t)$ and  
${\tilde B}^{(a)}(\zeta,\zeta';t,\lambda_t)$   
of the integral equations (3.58), depending meromorphically on the parameters $t$ and 
$\lambda_t$ for $(t,\lambda_t)$ varying in  
$\mathbb{R}^{-}\times \mathbb{C}_{+}^{(N_H)}.$

By construction, these functions 
${\tilde B}^{(s)}$ and  
${\tilde B}^{(a)}$ 
are interpolations in the $\lambda_t-$plane of the corresponding Euclidean partial waves 
$\{\tilde b_{2\ell}(\zeta,\zeta',t);\ 2\ell > N_H\}$ and 
$\{\tilde b_{2\ell+1}(\zeta,\zeta',t);\ 2\ell +1 >N_H\}$  
of the Bethe-Salpeter kernel $B([k]),$ namely there holds the following 
Froissart-Gribov-type equalities:
\begin{equation}
{\rm for}\ \ 2\ell >N_H,\ \ \  
{\tilde B}^{(s)}(\zeta,\zeta';t,2\ell) =   
\tilde b_{2\ell}(\zeta,\zeta',t), 
\tag{3.59}  
\end{equation}
\begin{equation}
{\rm for}\ \ 2\ell+1 >N_H,\ \ \  
{\tilde B}^{(a)}(\zeta,\zeta';t,2\ell+1) =   
\tilde b_{2\ell+1}(\zeta,\zeta',t), 
\tag{3.60}  
\end{equation}
According to the results of Proposition A3, in which $F$ is replaced by
$\tilde H^{(s)}$
(resp, $\tilde H^{(a)} $), 
we can make the following remark on the corresponding resolvent 
$R_{F|\alpha=-1} = \tilde B^{(s)}$ (resp. $\tilde B^{(a)}$). 

\vskip 0.2cm
\noindent 
{\bf Remark} \ \ the general bounds (3.57) on 
$\tilde H^{(s),(a)}(\zeta,\zeta';t,\lambda_t)$ lead one to a choice of the function 
$C(\lambda_t)$ of Proposition A3 proportional to   
$\left| \lambda _{t}-N_H\right|
^{n_H-\frac{d-2}{2}}$,   
which implies bounds of the type (A.22) on  
$\tilde  B^{(s),(a)}(\zeta,\zeta';t,\lambda_t) \times 
{\cal D}_{\tilde H^{(s)(a)}|\alpha= -1}(t,\lambda_t) $.   
These bounds are comparable to (3.57)
concerning their dependence  
in the mass variables, but contain in general no temperate dependence with respect to $\lambda_t$ in  
$\mathbb C_+^{(N_H)}$; this also corresponds to the impossibility of controlling a priori 
the poles of 
$\tilde  B^{(s),(a)}$ in the limit $|\lambda_t| \to \infty$.  

\vskip 0.2cm
In the following, we shall always be led to assume that $B$ satisfies an extra-assumption of temperateness 
which leads us to the following statement

\vskip 0.2cm
\noindent 
{\bf Proposition 1} \ \ \  {\sl If    
$\Delta^{(s)}{\underline B}_{(\zeta,\zeta',t)}(\cosh v)$ 
and $\Delta^{(a)}{\underline B}_{(\zeta,\zeta',t)}(\cosh v)$ 
have a behaviour at infinity which is governed by 
${\rm e}^{N_B v}$ (with $N_B \geq N_H$), then the solutions 
$\tilde B^{(s)}$ and 
$\tilde B^{(a)}$  of the BS-equation (3.58) are holomorphic functions of $\lambda_t$ 
in the half-plane 
$\mathbb C_+^{(N_B)}$,
and coincide there with the following $L_d-$transforms 
\begin{equation}
{\underline {\tilde B}^{(s),(a)}}(\zeta,\zeta';t,\lambda_t)=   
\omega_{d-1}\int_0^{\infty}
\Delta^{(s)(a)}{\underline B}_{(\zeta,\zeta',t)}(\cosh v) Q^{(d)}_{\lambda_t}(\cosh v) 
(\sinh v)^{d-2} dv 
\tag{3.61}  
\end{equation}
Moreover, the BS-equations (3.58) then appear as ``diagonalized forms'' of the original 
BS-equations (3.47),(3.48)}.

\vskip 0.3cm
For simplicity, we give the proof of this result for the case when 
$\Delta^{(s),(a)} \underline B_{(\zeta,\zeta',t)}$ are locally integrable functions. 
Applying the $L_d-$transformation (3.50) side-by-side to 
Eqs (3.47),(3.48), one obtains for $\lambda_t$ in 
$\mathbb C_+^{(N_B)}$ (in view of Eqs (3.54), (3.61)):
\begin{equation*}
\tilde{H}^{\left( s\right) ,\left( a\right) }\left( \zeta ,\zeta ^{\prime
};t,\lambda _{t}\right) =\underline {\tilde B}^{\left( s\right) ,\left( a\right) }\left(
\zeta ,\zeta ^{\prime };t,\lambda _{t}\right) +
\omega_{d-1}\int_0^{\infty}
Q^{(d)}_{\lambda_t}(\cosh v) 
(\sinh v)^{d-2}  \times \ldots  
\end{equation*}
\begin{equation}
\left[\int\limits_{\Delta _{t}}
\left( \Delta ^{\left( s\right) ,\left( a\right) }
{\underline{B}}_{\left( \zeta ,\zeta ^{\prime \prime },t\right) } \blacklozenge  
\Delta ^{\left( s\right) ,\left( a\right) }{\underline{H}}_ 
{(\zeta'' ,\zeta' ,t) }\right) \left( \cosh v\right)
\ \ \underline{G}\left( \zeta ^{\prime \prime }\right) 
d\mu _{t}\left( \zeta ^{\prime \prime }\right) 
\right] dv 
\tag{3.62}
\end{equation}
We shall apply the result of [12] (proposition III-2) according to 
which 
the $L_{d}-$transform  
of the convolution product $F_{1}\lozenge F_{2}$ of invariant
Volterra kernels $F_{i}\left( z,z^{\prime }\right) =f_{i}\left( -z \cdot  
z^{\prime }\right) ,i=1,2,$ namely $f_{1}\blacklozenge f_{2}$ (in view of our
definition (3.35)), is equal to the product $\tilde{F}_{1}\left( \lambda
\right) \times \tilde{F}_{2}\left( \lambda \right) .$ 

More precisely (see proposition II-2i) of [12]), if the kernels $F_i\ (i=1,2)$ 
are regular functions satisfying norm conditions of the form 
\begin{equation}
g_N(f_i)= \int_0^{\infty} {\rm e}^{-Nv}\ |f_i(\cosh v)| dv\  <\  \infty,   
\tag{3.63}
\end{equation}
then their Volterra convolution product satisfies a bound of the form  
\begin{equation}
g_N(f_1 \blacklozenge f_2) \leq c_N \  g_N(f_1) \ g_N(f_2)\  <\  \infty,  
\tag{3.64}
\end{equation}
where $c_N$ is a numerical constant. Correspondingly, 
the product ${\tilde F}_1 \times  
{\tilde F}_2$ is holomorphic in   
$\mathbb{C}^{(N)}_+$ and one has: 
\begin{equation*}
|{\tilde F}_1(\lambda)
{\tilde F}_2(\lambda)| \leq  
{\rm cst}\ \int_0^{\infty} |(f_1 \blacklozenge f_2)(\cosh v)| \ |Q_{\lambda}^{(d)}(\cosh v)| 
(\sinh v)^{d-2} dv 
\end{equation*}
\begin{equation}
\leq\  c_{(\varepsilon)}\ g_N(f_1 \blacklozenge f_2), 
\tag{3.65}
\end{equation}
uniformly in any half-plane 
$\mathbb{C}^{(N + \varepsilon)}_+$  
(for all $\varepsilon >0$ and a suitable choice of the constant 
$ c_{(\varepsilon)}$): this majorization follows from the exponential 
decrease property of the function 
$Q_{\lambda}^{(d)}(\cosh v)$ (see [12]). 

Assuming that the absorptive parts 
$\Delta^{(s),(a)} H,$
$\Delta^{(s),(a)} B$
satisfy norm conditions of the form (3.63) with the mass dependence given 
by Eq. (3.28), namely  
$g_{N_B}(\Delta^{(s),(a)}{\underline  H}_{(\zeta,\zeta',t)} ) 
= c_H \ C^{(H)}_{(\zeta,\zeta',t)} $
(as it is implied by (3.27) if $n_H =0$)
and $g_{N_B}(\Delta^{(s),(a)}{\underline  B}_{(\zeta,\zeta',t)} ) 
= c_B \ C_{(\zeta,\zeta',t)}$,   
($c_H$ and $c_B$ being numerical constants), then it follows from 
(3.64),(3.65) that the repeated integral 
$$\int\limits_{\Delta _{t}}
\left[
\int_0^{\infty}
\left( \Delta ^{\left( s\right) ,\left( a\right) }
{\underline{B}}_{\left( \zeta ,\zeta ^{\prime \prime },t\right) }\  \blacklozenge \  
\Delta ^{\left( s\right) ,\left( a\right) }{\underline{H}}_ 
{(\zeta'' ,\zeta' ,t) }\right) \left( \cosh v\right)
Q^{(d)}_{\lambda_t}(\cosh v) 
(\sinh v)^{d-2} dv \right] $$    
\begin{equation}
\ldots \times \underline{G}\left( \zeta ^{\prime \prime }\right) 
d\mu _{t}\left( \zeta ^{\prime \prime }\right) 
\tag{3.66}
\end{equation}
is absolutely convergent and therefore equal to the double integral at the r.h.s. of Eq.(3.62). 
The integral over $v$ in (3.66) is the $L_d-$transform of a Volterra convolution, 
equal to the product 
$\underline {\tilde{B}}^{\left( s\right) ,\left( a\right) }\left(
\zeta ,\zeta'' ;t,\lambda _{t}\right)\  
\tilde{H}^{\left( s\right) ,\left( a\right) }\left( \zeta'' ,\zeta' 
;t,\lambda _{t}\right)$ in
$\mathbb C_+^{(N_B)}.$
This shows that Eq.(3.62) coincides with the corresponding BS-equation (3.58) 
for the Laplace transforms, so that  
$\underline {\tilde{B}}^{\left( s\right) ,\left( a\right)}$ coincides with the restriction of 
${\tilde{B}}^{\left( s\right) ,\left( a\right)}$ to the half-plane   
$\mathbb C_+^{(N_B)}$. 

The case when 
$ \Delta ^{\left( s\right) ,\left( a\right) }
{\underline{B}}_{\left( \zeta ,\zeta ^{\prime \prime },t\right) }$ 
are distributions can be treated similarly, since the convolution product 
$F_1 \lozenge F_2$ of distribution-like invariant Volterra kernels is 
again transformed by $L_d$ in the corresponding product 
$\tilde F_1(\lambda) 
\times \tilde F_2(\lambda)$, these 
holomorphic functions admitting now a power bound in $|\lambda|$ 
instead of being bounded.  

\vfill\eject

\section{On the Bethe-Salpeter generation of Regge poles in general
quantum field theory}

In Sec 3 we have exhibited how the BS-type 
structure resulting from the general 
(axiomatic) framework of QFT can be expressed 
in terms of the squared-mass and angular variables, 
and then in terms of the squared-mass and 
{\sl complex angular momentum} variables; at each step, 
this was done by considering 
the kernel $B$ (resp. $\tilde B^{(s),(a)}$) 
as determined by $H$ via the Fredholm method. 
In the present section, we shall adopt a more 
exploratory viewpoint by assuming that in 
the field theory under consideration, the kernel 
$B([k]) = B[t; \rho,w,\rho',w'; \cos \Theta_t]$ satisfies 
``better properties''
than $H$, in a sense to be specified below.  We then intend to draw the
consequences of these additional assumptions on $B$ for the structure of
$H$.

\subsection{Local generation of Regge poles}

We consider again the diagonalized form (3.58) of the Bethe-Salpeter
structure of a given four-point function $H$, assuming that the
conditions
of Proposition 1 are satisfied  with  $N_B= N_H$. Eq. (3.58) is then
valid as an identity between  holomorphic functions of $\lambda_t$
in the half-plane $\mathbb C_+^{(N_H)}$, for all values of
$(\zeta,\zeta')$
in $\Delta_t \times \Delta_t$ and  negative $t$.  Moreover
$\tilde B^{(s),(a)}$ satisfy bounds which are similar to those on
$\tilde H^{(s),(a)}$, as far as their dependence on the mass
variables $\zeta,\zeta'$ is concerned, namely these bounds contain
(as in (3.57))  a factor  $C_{(\zeta,\zeta',t)}$ of the form (3.28).

Let us now make the additional assumption that $\tilde B^{(s),(a)}$
can be analytically continued in some disk of the $\lambda_t-$plane,
centered  on the border line ($\Re e\lambda_t = N_H$) of
$\mathbb C_+^{(N_H)}$ and still satisfy similar bounds including the
factor
$C_{(\zeta,\zeta',t)}$ in that disk. Let $D$ be the intersection of the
latter
with the closed left-hand plane $\Re e\lambda_t \leq N_H$;  we can
consider  the analytic continuation of  (3.58) in $D$, these equations
being now regarded as  Fredholm-resolvent  equations defining 
$\tilde
H^{(s)}$
(resp. $\tilde H^{(a)}$) in terms of  $\tilde B^{(s)}$ (resp. $\tilde
B^{(a)}$)
through expressions of the form:
\begin{equation}
\tilde H^{(s),(a)}(\zeta,\zeta';t,\lambda_t) = 
{{\cal N}_{\tilde B^{(s),(a)}}(\zeta,\zeta';t,\lambda_t) \over 
{\cal D}_{\tilde B^{(s),(a)}}(t,\lambda_t) }.   
\tag{4.1} 
\end{equation}
In Eq. (4.1), the notations of Proposition A3 have been simplified by
putting
$${\cal N}_{\tilde B^{(s),(a)}}(\zeta,\zeta';t,\lambda_t) = 
{\cal N}_{\tilde B^{(s),(a)}|\alpha =1}(t,\lambda_t; \rho,w,\rho',w') 
\quad {\rm and}$$ 
$${\cal D}_{\tilde B^{(s),(a)}}(t,\lambda_t) = 
{\cal D}_{\tilde B^{(s),(a)}|\alpha =1}(t,\lambda_t ). $$ 
Since $\tilde H^{(s)}$ and  $\tilde H^{(a)}$ are holomorphic in
${\mathbb C}_+^{(N_H)}$, the resolution (4.1) provides a meromorphic
continuation of  $\tilde H^{(s)}$ and $\tilde H^{(a)}$  in $D$, whose
poles
are given respectively by the zeros of 
${\cal D}_{\tilde B^{(s)}(t,\lambda_t) }$ and    
${\cal D}_{\tilde B^{(a)}(t,\lambda_t) }:$   
the locations of
the latter in the
$\lambda_t-$plane  will be denoted respectively by
$\lambda_t = \lambda_{j^{(s)}}(t), 
\lambda_t = \lambda_{j^{(a)}}(t)$.  
\vskip 0.2cm  
\noindent
{\bf Remark}   The variable $t$ being kept real (and negative) at the
present step of our program,  there is no point of proving  the analytic
dependence of the previous  zeros with respect to $t$;  however, this
analyticity will naturally appear at a further step, once one has
extended
the results of [1] to a relevant set of complex values of $t$ by
techniques
of analytic completion.

\vskip 0.2cm

For simplicity, we shall only consider the case when 
there is a 
unique zero
$\lambda^{(s)}(t)$ (resp. 
$\lambda^{(a)}(t)$) of  
${\cal D}_{\tilde B^{(s)}(t,\lambda_t) }$ (resp.    
${\cal D}_{\tilde B^{(a)}(t,\lambda_t) }$):   
in $D$, which we suppose to be a 
{\sl simple} zero; correspondingly, $\tilde H^{(s)}$
and $\tilde H^{(a)}$ admit these values $\lambda^{(s)}(t)$,
$\lambda^{(a)}(t)$  as simple poles, which can be called
{\sl Regge poles}, and we introduce the corresponding  residue
functions:
$${\rm Res}\ \tilde H^{(s),(a)}(\zeta,\zeta';t, \lambda^{(s),(a)}(t))$$ 
\begin{equation}
=  [(\lambda_t -\lambda^{(s),(a)}(t)) \times 
\tilde H^{(s),(a)}(\zeta,\zeta'; t,
\lambda_t)]
_{|\lambda_t = \lambda^{(s),(a)}(t)}. \tag{4.2} 
\end{equation}
\noindent
{\bf Factorization property of the Regge pole residues:}

\vskip 0.2cm
We claim that, as a result of Fredholm theory,  
the functions 
${\rm Res}\ \tilde H^{(s)}$ and  
${\rm Res}\  \tilde H^{(a)}$ can be written under the following form: 
\begin{equation}
{\rm Res}\ \tilde H^{(s)}(\zeta,\zeta'; t,\lambda^{(s)}(t))    
= {p^{(s)}(\zeta,\zeta',t) \over \beta^{(s)}(t)}, 
\tag{4.3}
\end{equation}
\begin{equation}
{\rm Res}\ \tilde H^{(a)}(\zeta,\zeta'; t,\lambda^{(a)}(t)) 
= {p^{(a)}(\zeta,\zeta',t) \over \beta^{(a)}(t)}, 
\tag{4.4}
\end{equation}
where $p^{(s)}$ and $p^{(a)}$ are (for each $t$) {\sl kernels of finite rank} 
satisfying the {\sl projector relations}  
$p^{(s)}\circ_{t} p^{(s)} = p^{(s)},$ 
$p^{(a)}\circ_{t} p^{(a)} = p^{(a)}.$  
This can be seen as follows. 

Making use of the expression (A.16) of the Fredholm resolvent 
$R_{\tilde B^{(s)}}$ of 
$\tilde B^{(s)}$, and calling $\alpha= \underline \alpha(t,\lambda_t)$ the 
zero of 
${\cal D}_{\tilde B^{(s)}}(t, \lambda_t;\alpha)$  
whose restriction to $\alpha =1$ yields the pole
$\lambda_t = \lambda^{(s)}(t)$ of $\tilde H^{(s)}$ 
($\lambda_t$ being thus a solution of the implicit equation 
$\underline \alpha(t,\lambda_t)-1 =0$), 
one can show (see e.g. Sec 3.3 of [17] 
and references therein to standard results of Fredholm theory)
that 
$$R_{\tilde B^{(s)}}(\zeta,\zeta'; t,\lambda_t,\alpha) = 
{p^{(s)}(\zeta,\zeta',t) \over  
\alpha - \underline \alpha(t,\lambda_t)}   
+R'_{\tilde B^{(s)}}(\zeta,\zeta'; t,\lambda_t,\alpha),$$ 
where $p^{(s)}$ is of finite rank and   
$R'_{\tilde B^{(s)}}$  
holomorphic in $\alpha$ at 
$\alpha= \underline \alpha(t,\lambda_t)$.
Formula (4.3) then follows by putting 
$ \beta^{(s)}(t) =  
-{\partial \underline \alpha \over \partial \lambda_t}(t,\lambda^{(s)}(t))$.  
The analysis would be more involved in the case of 
multiple poles (not considered here).   

\vskip 0.3cm
By assumption, we shall consider as ``generic'' 
the case when the Regge pole terms 
are characterized by {\sl projectors of rank one}; 
for the case of Hermitian fields $\phi_1$, $\phi_2$ considered here, 
these terms can always be 
written under the following form: 
\begin{equation}
p^{(s)}(\zeta,\zeta',t) = 
{\overline \varphi^{(s)}(\zeta,t)} 
\times \varphi^{(s)}(\zeta',t),  
\tag{4.5}
\end{equation}
\begin{equation}
p^{(a)}(\zeta,\zeta',t) = 
{\overline \varphi^{(a)}(\zeta,t)} 
\times \varphi^{(a)}(\zeta',t),  
\tag{4.6}
\end{equation}
with $\int_{\Delta_t} 
|\varphi^{(s)}(\zeta,t)|^2 d\mu_t(\zeta)  
= \int_{\Delta_t} 
|\varphi^{(a)}(\zeta,t)|^2 d\mu_t(\zeta) = 1$.
The Hermitian character of the r.h.s. of Eqs (4.5), (4.6) is in fact 
implied by the symmetry property 
$ \tilde H^{(s),(a)}(\zeta,\zeta'; t,\lambda_t)    
= \overline  {\tilde H^{(s),(a)}(\zeta,\zeta'; t,{\overline\lambda_t}) }$ 
together with the assumption that  $\lambda^{(s),(a)}(t)$ are real. 
To be more precise, if $H([k])$ is the four-point function of two 
{\sl Hermitian} scalar fields $\phi_1,\phi_2,$ it satisfies in its 
analyticity domain the symmetry relation
$H(K;Z,Z') = \overline{H(\overline K; \overline Z', \overline Z)}$. 
The latter implies the following ones (for real values of $t,\zeta,\zeta'$):
$\underline H_{(\zeta,\zeta',t)}(\cos \Theta_t)=  
\overline{\underline H_{(\zeta,\zeta',t)}(\overline {\cos \Theta_t})}$ 
and therefore also (in view of Eq. (3.54)):\  
$\tilde H^{(s),(a)} (\zeta,\zeta';t,\lambda_t) 
= \overline{\tilde H^{(s),(a)} (\zeta,\zeta';t,\overline{\lambda_t})}$. 
It follows that the poles 
$\lambda_t = \lambda_{j^{(s),(a)}}(t)$ resulting from Eq. (4.1)  
can only be produced either at real values or at pairs of 
complex conjugate values. 
However in the range $t<0$ considered here, 
the case of {\sl real} poles 
appears to be more ``physical'', 
as it can be illustrated by simple models of BS-kernels, and this 
justifies our reality assumption on $\lambda^{(s),(a)}(t)$.   

\vskip 0.2cm
\noindent
{\bf Remark}  The functions $\varphi^{(s),(a)}(\zeta,t)$ 
which depend on the  
three Lorentz invariants $\zeta_1,\zeta_2, t$ can be interpreted (from
the
geometrical viewpoint) as {\sl three-point functions} defined
in the Euclidean domain of
field theory, namely in  the set of configurations $(k_1,k_2,
k_3=k_1+k_2)$
such that $k_i = (iq_i^{(0)}, \vec p_i),\  i=1,2,3,$
whose invariants $\zeta_1= k_1^2, \  \zeta_2 = k_2^2 $ and  
$t=(k_1+k_2)^2$
vary in the set $\{(\zeta,t);\  \zeta \in \Delta(t),\  t <0 \}$.

\subsection{Asymptotic assumption on the Bethe-Salpeter\break  
kernel and generation of
dominant
Reggeon terms\break in the four-point functions:}

In this subsection, we shall assume that  
$B([k]) = B[t; \rho,w,\rho',w';\cos \Theta_t]$
satisfies increase
properties in the variable 
$\cos \Theta_t = -z\cdot z'$
which are {\sl ``better than''} those of $H$.  We then intend to show
that,  for any value of
$t$ for which this assumption is made,  
there exists a possible Regge
pole structure
of $\tilde H^{(s)} $  and $\tilde H^{(a)}$ which will produce
corresponding
asymptotically dominant ``Reggeon terms'' in the four-point function
$H([k])$.
For the negative values of the channel variable $t$
only considered  in the present paper,  it appears
difficult to give a
general field-theoretical support to this type of assumption, although
it is naturally satisfied
by standard approximations of the Bethe-Salpeter kernel $B$ such as the
relativistic
counterpart of Yukawa-type potentials.  As a matter of fact, it is for
$t$ varying in some
{\sl positive} interval that we shall be able to produce more general
arguments in favour of such
assumption on $B$, but this can only be relevant 
in our program after the 
analytic continuation
in the variable $t$ has been performed (in a further work). However,
since the
mathematical analysis is independent of the value of $t$ (playing the
role of a fixed parameter),
we think it appropriate to present it below as {\sl ``a generic
procedure for the generation
of asymptotically dominant Reggeon terms 
in the field-theoretical framework''}.

\vskip 0.2cm
More specifically, we shall now assume that 
$B$ satisfies 
the following bounds involving a given exponent $N_B $ and a positive constant $C_E^{(B)}$:

\quad\quad \quad\quad\quad  {\sl  i)}  
\ for all $u$\ \  
($(0\leq u\leq \pi\  {\rm and} \ \pi \leq u \leq 2\pi,$ the jumps at $u=0$ 
and $u=\pi$ being taken into account),  
\begin{equation}
\int_0^{\infty}  {\rm e}^{-N_B v}  
|\underline B_{(\zeta,\zeta',t)}(\cos (u+iv))| dv  \leq 
C^{(B)}_{(\zeta,\zeta',t)},  
\tag{4.7} 
\end{equation}
\begin{equation}
ii)\quad\quad\quad  \sup_{-1\leq\cos \Theta_t \leq 1}  
|\underline B_{(\zeta,\zeta',t)}(\cos \Theta_t)|  \leq 
C^{(B)}_{(\zeta,\zeta',t)},  
\tag{4.8} 
\end{equation}
with  
\begin{equation}
C^{(B)}_{(\zeta,\zeta',t)}=   
C_E^{(B)} [(1+|t|^{1\over 2}) \ (1+\rho)\ (1+ |w||t|^{1\over 2}) 
 \ (1+\rho')\ (1+ |w'||t|^{1\over 2})]^{N_B}  . 
\tag{4.9} 
\end{equation}
In these assumptions, {\sl the real number $N_B$ is supposed to satisfy the conditions} 
$\max (-1,-{d-2\over 2}) < N_B < N_H$. So, if we compare the previous 
assumptions on $B$ with the corresponding bounds 
(3.27), (3.28) on $H$, we see that they differ under two respects: 

a) the increase at infinity in the $\cos \Theta_t-$plane is governed by 
${\rm e}^{N_B |\Im m \Theta_t|}$ (dominated by  
${\rm e}^{N_H |\Im m \Theta_t|}$),  

b) the use of an $L^1-$bound instead of a uniform bound (as for $H$ in (3.27))
is motivated by its convenience for the $\star^{(c)}-$convolution formalism on 
$X_d^{(c)}$. Here, the local order of singularity of the boundary values and
discontinuities $\Delta^{(s),(a)} B$ of $B$ is encoded in their $L^1-$character with respect to 
the variable $v= \Im m \Theta_t$ together with their uniform dependence on 
$\sin \Re e \Theta_t$.

We then have: 

\vskip 0.2cm
\noindent
{\bf Proposition 2}\  
{\sl The previous assumptions i), ii) on $B$ imply the 
analyticity property 
with respect to $\lambda_t$ in ${\mathbb C_+^{(N_B)}}$ 
and the following majorizations for the transforms 
${\tilde B}^{(s)}$, 
${\tilde B}^{(a)}$ of 
$\Delta^{(s)} B$,
$\Delta^{(a)} B$:
\begin{equation}
|{\tilde B}^{(s),(a)}(\zeta,\zeta';t,\lambda_t)| \leq   
C^{(B)}_{(\zeta,\zeta',t)} 
\Psi(|\Im m \lambda_t|),  
\tag{4.10} 
\end{equation}
where $\Psi$ denotes a bounded positive function on $[0,\infty[$, tending to zero at infinity.  
These majorizations hold uniformly for all $(\zeta,\zeta') \in \Delta_t \times \Delta_t$,
$t<0$ and $\lambda_t \in {\mathbb C_+^{(N_B)}}$. } 

\vskip 0.3cm

The proof of the latter relies on the fact that the action of the 
$L_d-$transformation on 
$\Delta^{(s),(a)} B$
factorizes as follows 
(see [12], formulae (III-1),(III-3)): 
\begin{equation}
{\tilde B}^{(s),(a)}(\zeta,\zeta';t,\lambda_t)= \int_{\underline v}^{\infty} {\rm e}^{-\lambda_t w}
\widehat{\Delta^{(s),(a)} B}(\zeta,\zeta';t,w) dw, 
\tag{4.11} 
\end{equation}
with
\vfill \eject

$$\widehat{\Delta^{(s),(a)} B}(\zeta,\zeta';t,w) = \cdots $$ 
\begin{equation}
\omega_{d-2} {\rm e}^{-{d-2\over 2}}
\int_0^w 
\Delta^{(s),(a)}{\underline  B}_{(\zeta,\zeta',t)}(\cosh v)  
[2(\cosh w-\cosh v)]^{d-4\over 2} \sinh v  dv.  
\tag{4.12} 
\end{equation}
It can be seen (see corollary of Proposition II-6 in [12]) that 
in view of the condition  $N_B > -1$
the $L^1-$bound on 
$\Delta^{(s),(a)}{\underline  B}_{(\zeta,\zeta',t)}(\cosh v)$ 
deduced from (4.7), namely
\begin{equation}
\int_0^{\infty} {\rm e}^{- N_B v} \     
|\Delta^{(s),(a)}{\underline  B}_{(\zeta,\zeta',t)}(\cosh v)|\ dv  
\leq C_{(\zeta,\zeta', t)}^{(B)}\tag {4.13} 
\end{equation}
implies the same 
$L^1-$bound (up to a constant factor) on  
$\widehat{\Delta^{(s),(a)} B}(\zeta,\zeta';t,w)$
Then since Eq. (4.11) represents a usual Fourier-Laplace transformation, 
the announced bound (4.10) readily 
follows from the Lebesgue theorem.

\vskip 0.3cm
\noindent
{\bf Remark} \ \  a similar result holds if the bound (4.13) is satisfied 
only in the sense of distributions: this is the case e.g. if 
$B$ contains a simple or multiple pole of the form 
$(\cos \Theta_t - \cosh v_0)^{-q}$, interpreted as a 
``particle-exhange''
(or Yukawa-type) contribution for $q=1$, or as a 
more singular ``gluon-type exchange'' contribution for $q > 1$. 

\vskip 0.3cm
in view of Proposition 2, the considerations of subsection 4.1 apply
to the analytic continuations of the kernels 
$\tilde B^{(s)}$ and $\tilde B^{(a)}$ in the full half-plane 
$ \Re e\lambda_t > N_B$. It follows that  
$\tilde H^{(s)}$ and $\tilde H^{(a)}$ 
admit meromorphic continuations of the form (4.1) in the strip
$N_B < \Re e\lambda_t \leq N_H$. In this strip, there will occur
possible Regge poles equipped with factorized residues of the form
described by Eqs (4.3),..,(4.6).
Moreover we notice that, in view of the bounds (4.10) on  
$\tilde B^{(s),(a)}$, the last statement of Proposition A3 
can be applied. It follows that,
under our assumptions on $B$, the poles  
$\lambda_t = \lambda_{j^{(s)}}(t), 
\lambda_t = \lambda_{j^{(a)}}(t)$ of   
$\tilde H^{(s)}$ and
$\tilde H^{(a)}$ are confined in a bounded region
of the form  
$\{\lambda_t; \ N_B< \Re e \lambda_t \leq N_H ,
\ |\Im m\lambda_t|< \nu_{\Psi}(t)\}$.  
Finally, by applying the majorization (A.22) with 
$C(\lambda_t) = C^{(B)}_E 
\Psi(|\Im m \lambda_t|)$, 
we obtain bounds of the following form which hold uniformly for all 
$(\zeta,\zeta')\in \Delta_t \times \Delta_t$, 
$N_B < \Re e \lambda_t \leq N_H,$ 
(and $t \leq - \varepsilon$, for any given  
$\varepsilon$,  
$\varepsilon >0$):   
\begin{equation}
|{\cal D}_{B^{(s),(a)}}(t, \lambda_t) \times  
\tilde H^{(s),(a)}(\zeta,\zeta';t,\lambda_t)|  
\leq 
\hat C^{(B)}_{(\zeta,\zeta',t)} 
\ \Psi(|\Im m \lambda_t|),  
\tag{4.14} 
\end{equation}
where
$\hat C^{(B)}_{(\zeta,\zeta',t)}$
is given by Eq. (A.22).
Besides, for $|\Im m \lambda_t| > A(t)$, with $A(t)$ sufficiently large, and 
$N_B < \Re e \lambda_t \leq N_H,$ 
one can satisfy the inequality
$|{\cal D}_{B^{(s),(a)}}(t, \lambda_t) -1| < {1\over 2}$ (in view of (A.14)), and 
therefore the bound (4.14) can be advantageously replaced by 
\begin{equation}
|\tilde H^{(s),(a)}(\zeta,\zeta';t,\lambda_t)|  
\leq 2 
\hat C^{(B)}_{(\zeta,\zeta',t)} 
\Psi(|\Im m \lambda_t|).  
\tag{4.15} 
\end{equation}

\vfill \eject
\noindent
{\bf Reggeon structure of the four-point function:}  

\vskip 0.3cm
We shall now show that the previous properties of meromorphic continuation of 
$\tilde H^{(s)}$ 
and $\tilde H^{(a)}$ 
imply the existence of a peculiar structure of the four-point function 
$H([k]) \equiv 
\underline H_{(\zeta,\zeta',t)}(\cos \Theta_t) $ 
as a function of $t$ and $\cos \Theta_t$. 
This will be done by starting from the inversion formula (see Eq (4.89) in Theorem 5 of [1]), 
which allows one to reexpress  
$\underline H_{(\zeta,\zeta',t)}(\cos \Theta_t) $ 
{\sl for all $\cos \Theta_t$ in the cut-plane} 
$\Pi = {\mathbb C}\setminus \{[\cosh v_0, +\infty[ 
\ \cup\  ]-\infty,-\cosh v_0]\}$ 
in terms of 
$\tilde H_s= {\tilde H^{(s)} + \tilde H^{(a)}\over 2}$
and $\tilde H_u= {\tilde H^{(s)} - \tilde H^{(a)}\over 2} $. 
In view of the bounds (3.27) on $H$ and correspondingly of the bounds (3.57) 
on $ \tilde H^{(s)}$,$ \tilde H^{(a)}$
in ${\mathbb C}_+^{(N_H)}$, formula (4.89) of [1] can be applied 
\footnote{The choice of the line $\Re e \lambda = N_H + \varepsilon$
($\varepsilon$ arbitrarily small) for the integration cycle in Eq. (4.15) 
is more correct than the prescription ($\Re e \lambda = N_H$) 
given in Eq. (4.89) of [1], which
necessitates uniform bounds in the closure of
${\mathbb C}_+^{(N_H)}$}  
and yields:  

$$\underline H_{(\zeta,\zeta',t)}(\cos \Theta_t) = $$ 
$$-{1\over 2i \omega_d} \int_{N_H +\varepsilon -i\infty} 
^{N_H +\varepsilon +i\infty} 
\tilde H^{(s)}(\zeta,\zeta';t,\lambda)\  {{\rm h}_d(\lambda)  
[P_{\lambda}^{(d)}(\cos(\Theta_t-\varepsilon_{\Theta_t} \pi) + 
P_{\lambda}^{(d)}(\cos(\Theta_t)] \over 2\sin \pi \lambda} d\lambda $$  
$$-{1\over 2i \omega_d} \int_{N_H +\varepsilon -i\infty} 
^{N_H +\varepsilon +i\infty} 
\tilde H^{(a)}(\zeta,\zeta';t,\lambda)\  {{\rm h}_d(\lambda)  
[P_{\lambda}^{(d)}(\cos(\Theta_t-\varepsilon_{\Theta_t} \pi) - 
P_{\lambda}^{(d)}(\cos(\Theta_t)] \over 2\sin \pi \lambda} d\lambda $$  
\begin{equation}
+ {1\over \omega_d} \sum_{0\leq {\ell} \leq N_H} 
\tilde h_{\ell}(\zeta,\zeta',t) 
\ {\rm h}_d ({\ell}) P_{\ell}^{(d)}(\cos \Theta_t).  
\tag{4.16}
\end{equation}

In the latter, ${\rm h}_d(\lambda) = {2\lambda +d-2 \over (d-2)!} \times 
{\Gamma(\lambda + d-2) \over \Gamma(\lambda +1)}$, $\varepsilon (\Theta_t) 
= {\rm sgn}(\Re e \Theta_t)$ and $ P_{\lambda}^{(d)}(\cos \Theta_t)$ 
denotes the first-kind Legendre function in dimension $d$ , whose 
integral representation is given by Eq. (4.64) of [1]. 

The absolute convergence of the integral at the r.h.s. of Eq. (4.16) is 
ensured for all $\cos \Theta_t$ in $\Pi$ ( uniformly in 
$\cos \Theta_t$) by appropriate exponential decrease properties of the 
integration kernels with respect to the variable $\nu = \Im m \lambda$. 
These properties result from the following bounds on the Legendre 
functions 
$ P_{\lambda}^{(d)}$ (obtained e.g. from formulae (II.85), (II.86) of
[12] by a refinement of the argument of Proposition II-12):   
\begin{equation}
\left|{{\rm h}_d(N+ i\nu)
P_{N+ i\nu}^{(d)}(\cos \Theta_t)\over \sin \pi (N+i\nu)}\right| \leq  
C_d(\cos \Theta_t) {\rm e}^{\max (N, -N-d+2)|\Im m\Theta_t|} 
{\rm e}^{-|\nu|(\pi -|\Re e \Theta_t|)},  
\tag{4.17}
\end{equation}
where the function
$C_d(\cos \Theta_t)$ is uniformly bounded at infinity in the cut-plane 
${\mathbb C} \setminus ]-\infty,-1]$. 

By now taking advantage of the meromorphic continuation of 
$\tilde H^{(s)}$ 
and $\tilde H^{(a)}$ in the strip $N_B < \Re e \lambda_t \leq N_H$ 
together with the uniform bound (4.15) on these functions when 
$|\Im m \lambda_t|$ tends to infinity, one can shift the integration line 
in formula (4.16) from its 
initial position to the line $\Re e \lambda_t = N_B$,
provided one extracts residue terms corresponding to all the poles 
$\lambda_t = \lambda_{j^{(s)}}(t), 
\lambda_t = \lambda_{j^{(a)}}(t)$ 
contained in the strip. Assuming there exists 
only {\sl one simple pole} at 
$\lambda_t = \lambda^{(s)}(t)$ real for $\tilde H^{(s)}$  
and {\sl one simple pole} at  
$\lambda_t = \lambda^{(a)}(t)$ real for $\tilde H^{(a)}$, 
the expression (4.16) of 
$\underline H_{(\zeta,\zeta',t)}(\cos \Theta_t) $ 
can be replaced by:  
$$\underline H_{(\zeta,\zeta',t)}(\cos \Theta_t) = $$ 
$$-{\pi\over \omega_d} 
{\rm Res}\ \tilde H^{(s)}(\zeta,\zeta'; t,\lambda^{(s)}(t))    
{{\rm h}_d(\lambda^{(s)}(t)) \   
[P_{\lambda^{(s)}(t)}^{(d)}(\cos(\Theta_t-\varepsilon_{\Theta_t} \pi) + 
P_{\lambda^{(s)}(t)}^{(d)}(\cos(\Theta_t)] \over 2\sin \pi \lambda^{(s)}(t)}  $$  
$$-{\pi\over \omega_d} 
{\rm Res}\ \tilde H^{(a)}(\zeta,\zeta'; t,\lambda^{(a)}(t)) 
{{\rm h}_d(\lambda^{(a)}(t)) \   
[P_{\lambda^{(a)}(t)}^{(d)}(\cos(\Theta_t-\varepsilon_{\Theta_t} \pi) - 
P_{\lambda^{(a)}(t)}^{(d)}(\cos(\Theta_t)] \over 2\sin \pi \lambda^{(a)}(t)}  $$  
$$-{1\over 2i \omega_d} \int_{N_B -i\infty} 
^{N_B +i\infty} 
\tilde H^{(s)}(\zeta,\zeta';t,\lambda)\  {{\rm h}_d(\lambda)  
[P_{\lambda}^{(d)}(\cos(\Theta_t-\varepsilon_{\Theta_t} \pi) + 
P_{\lambda}^{(d)}(\cos(\Theta_t)] \over 2\sin \pi \lambda} d\lambda $$  
$$-{1\over 2i \omega_d} \int_{N_B -i\infty} 
^{N_B +i\infty} 
\tilde H^{(a)}(\zeta,\zeta';t,\lambda)\  {{\rm h}_d(\lambda)  
[P_{\lambda}^{(d)}(\cos(\Theta_t-\varepsilon_{\Theta_t} \pi) - 
P_{\lambda}^{(d)}(\cos(\Theta_t)] \over 2\sin \pi \lambda} d\lambda $$  
\begin{equation}
+ {1\over \omega_d} \sum_{0\leq {\ell} \leq N_B} 
\tilde h_{\ell}(\zeta,\zeta',t) \ {\rm h}_d ({\ell}) P_{\ell}^{(d)}(\cos \Theta_t).  
\tag{4.18}
\end{equation}
In the latter, the functions 
${\rm Res}\ \tilde H^{(s),(a)}$ 
are of the form given by Eqs (4.2),...,(4.6).
The important feature of Eq. (4.18) concerns the asymptotic
behaviour in the $\cos \Theta_t-$plane of the various terms at 
its right-hand side. 
In view of the 
dependence on $\Im m \Theta_t$ of the bound (4.17), it follows that 
the two integrals as well as the last term at the r.h.s. of Eq. (4.18) 
are functions of $\cos \Theta_t$ which are bounded at infinity by 
$|\cos \Theta_t|^{N_B}$. The first two terms at the r.h.s. of Eq. (4.18) 
then appear as the leading terms giving the asymptotic behaviour at 
infinity of
$\underline H_{(\zeta,\zeta',t)}(\cos \Theta_t) $. 
This asymptotic behaviour exhibits rates of increase which are those of 
$P_{\lambda^{(s)}}^{(d)} (\pm \cos \Theta_t)$
and $P_{\lambda^{(a)}}^{(d)} (\pm \cos \Theta_t)$, namely respectively 
$|\cos \theta_t|^{\lambda^{(s)}(t)}$
and $|\cos \theta_t|^{\lambda^{(a)}(t)}$,  
with $N_B < \lambda^{(s),(a)}(t) \leq N_H$. 
These leading terms in Eq. (4.18) will be called {\sl Reggeon terms}.   
To summarize, we can state:
\vskip 0.1cm
\noindent
{\bf Theorem}

{\sl Let
$H([k]) = \underline H_{(\zeta,\zeta',t)}(\cos \Theta_t)$  
be the analytic four-point function of two (mutually) 
local Hermitian scalar fields 
whose Bethe-Salpeter structure in the $t-$channel is encoded in a 
BS-kernel $B([k])= \underline B_{(\zeta,\zeta',t)}(\cos \Theta_t)$. 
Let us assume that for some range of negative values of $t $
(or possibly the whole half-line $t <0$),   
$B$ satisfies bounds of the form 
(4.7),(4.8),(4.9).  
Then in a generic situation involving only one Reggeon term of each 
symmetry type, 
$\underline H_{(\zeta,\zeta',t)}$ admits a representation of the 
following form, valid for all $(\zeta,\zeta')$ in 
$\Delta_t \times \Delta_t$ and $\cos \Theta_t$ in $ \Pi$:  
$$\underline H_{(\zeta,\zeta',t)}(\cos \Theta_t) = $$ 
$$-{\pi\over \omega_d} 
{{\overline \varphi^{(s)}(\zeta,t)} 
\times \varphi^{(s)}(\zeta',t) \over \beta^{(s)}(t)}   
{{\rm h}_d(\lambda^{(s)}(t)) \   
[P_{\lambda^{(s)}(t)}^{(d)}(\cos(\Theta_t-\varepsilon_{\Theta_t} \pi) + 
P_{\lambda^{(s)}(t)}^{(d)}(\cos(\Theta_t)] \over 2\sin \pi \lambda^{(s)}(t)}  $$  
$$-{\pi\over \omega_d} 
{{\overline \varphi^{(a)}(\zeta,t)} 
\times \varphi^{(a)}(\zeta',t) \over \beta^{(a)}(t)}  
{{\rm h}_d(\lambda^{(a)}(t)) \   
[P_{\lambda^{(a)}(t)}^{(d)}(\cos(\Theta_t-\varepsilon_{\Theta_t} \pi) - 
P_{\lambda^{(a)}(t)}^{(d)}(\cos(\Theta_t)] \over 2\sin \pi \lambda^{(a)}(t)}  $$  
\begin{equation}
+ \underline H'_{(\zeta,\zeta',t)}(\cos \Theta_t),  
\tag{4.19}
\end{equation}
where the residual term
$ \underline H'_{(\zeta,\zeta',t)}(\cos \Theta_t)$ is bounded at infinity  
by Cst$|\cos \Theta_t|^{N_B}$ in $\Pi$,  
while the first two terms exhibit leading behaviours 
governed respectively by  
$|\cos \Theta_t|^{\lambda^{(s)}(t)}$
and $|\cos \Theta_t|^{\lambda^{(a)}(t)},$
with $\lambda^{(s)}(t) > N_B $
and $\lambda^{(a)}(t) > N_B.$} 

\vskip 0.2cm
\section{Conclusion}

In this paper,  we have established the following results:

a) the general Bethe-Salpeter structure of four-point functions of
scalar fields
in a given $t-$channel has been shown to be diagonalized in the
corresponding
complex angular momentum variable $\lambda_t$  for all negative values
of  $t$
and of the squared-mass variables corresponding to Euclidean
configurations
in complex momentum space .  More specifically,  Carlsonian
interpolations
of the Euclidean Bethe-Salpeter equations for even and odd partial
waves have been constructed  in a half-plane of the form
${\rm Re}\  \lambda_t > N$,
starting from the corresponding Bethe-Salpeter equations for the $s$ and
$u-$channel absorptive parts.

b) any information on
$B$ leading to continue analytically
these interpolations in some region of the left-hand
half-plane ${\rm Re}\  \lambda_t \leq N$
results in the ``potential generation'' of Regge poles,  
equipped with a factorized residue structure
involving Euclidean three-point functions.

c) the existence of a meromorphic continuation of these
interpolations
in a strip of the form $N_b < {\rm Re}\ \lambda_t \leq  N$ 
(with the relevant
factorization
property of the residues of the poles) has been shown to hold if and
only if  the
Bethe-Salpeter kernel  satisfies bounds of the form
 $| \cos \Theta_t|^ { N_B}$ in the complex plane of the 
off-shell  scattering angle  
$\Theta_t$, which are better than those on the complete
four-point function $H$, namely iff  $N_B < N_H$. 

\vskip 0.1cm
Here our analysis has been done for
negative
values of $t$ by relying on the analyticity domains resulting from the
general principles of  quantum field theory.  It is expected to hold
similarly for
$t$ in some positive interval $]0, t_0[$,
and then to produce the desired interpolation of possible
bound states in the $t-$channel,
after one has justified the necessary properties
of analytic continuation of $H$ (in the
spirit of  [15]).

\vskip 0.1cm

In the general arguments of our axiomatic approach, the unspecified
number $N_H$
was introduced (see [1]) as a consequence of the temperateness
assumption in the
field-theoretical framework.  In a specific model of field theory, the
actual asymptotic behaviour of $H$ in the $\cos \Theta_t-$plane
would be governed by a precise function
$N_H =N_H(t)$.  Then  the crucial question of the validity
of Reggeon terms with asymptotic dominance
in the four-point function $H$
of a general quantum field theory  appears to
rely
basically on  the knowledge of the exponent $N_B = N_B(t)$ which governs
the asymptotic behaviour of $B$: such
Reggeon terms will appear for all values of
$t$
such that $N_B(t) <  N_H(t)$.
\vskip 0.1cm

More questionable is the justification of the inequality
$N_B(t) < N_H(t)$, which would require further investigations in the
present field-theoretical framework.  In the absence of global
information
on  general (regularized or renormalized) versions of the two-particle
irreducible
kernel $B$ , one can think of exploiting various contributions to $B$
defined
in terms of  generalized Feynman convolutions enjoying the graph
property of
``two-particle irreducibility''.  The simplest ones are of course
those associated with ``particle-exchange''
graphs, which are interpreted as  relativistic Yukawa-type potentials;
they do
satisfy the previous inequality (with $N_B= -1$) and have been widely
exploited
in the literature, since they directly transfer to particle physics the
initial procedure of
Regge pole production in potential theory.  However, much more
complicated
contributions to $B$ (of higher-order in the coupling constants)
are imposed by the Feynman convolution structure of field
theory; typical examples of those have been studied by Mandelstam [18]
and others
(see e.g. [19] and references therein). The
treatment of such contributions
which these authors have given  by using the methods  of  (maximally
analytic)
$S-$matrix theory, has led them to introduce ``Regge cuts'' in the
$\lambda_t-$plane,
whose effect in some situations seems to wipe out the dominant
asymptotic
behaviour of  the potential-type Regge poles introduced at first.
Although no mechanism of production of Regge cuts appears in
our general
field-theoretical framework, it can be seen in our approach  that 
the
contributions considered by the previous authors cannot satisfy 
the inequality $N_B(t) < N_H(t)$ 
for negative $t$, although they hopefully do it for
positive
$t$. Of course, such results on partial contributions to $B$
do not allow one
to conclude  
the non-validity of the Bethe-Salpeter generation of
Reggeon leading terms at negative $t$; more global
(non-perturbative) information on the field models 
would be needed for investigating the relevant asymptotic
properties of $B$.

\centerline{\bf APPENDIX }

\vskip 0.5cm
We give a unified treatment of the various versions of BS-equation  presented in Sec.3., 
in which the integration space either contains or is identical with the space of 
mass variables $\{\zeta =(\zeta_1,\zeta_2)\in \Delta_t \}$ equivalent to
$\{(\rho, w); \rho \geq 0, w\in {\mathbb R}\}$ (see Eqs (3.3)...(3.6)). One is led to use the 
${\cal N / \cal D} -$resolution of Fredholm resolvent equations in a complex analysis framework, in 
the following way.

$F(t,\star; \rho,w,\rho',w',\bullet,\bullet')$
will denote a $t-$dependent kernel acting on the integration space  
${\cal I}=\{(\rho, w,\bullet); \rho \geq 0, w\in {\mathbb R} , (\bullet \in \Gamma)\}$ 
($((\rho, w,\bullet), 
(\rho', w', \bullet')) \in  
{\cal I}\times  
{\cal I}$).   
The notation $\bullet \ (\bullet')$ stands for possible additional variables $z\ (z')$ which 
vary on a $(d-1)-$ cycle with compact support $\Gamma$ of 
$X^{(c)}_{d-1}$: this is the case for the kernel  
$H[t; \rho,w,\rho',w'; -z\cdot z'] \equiv \underline H_{(\zeta,\zeta',t)}(-z\cdot z')$  
(resp. $B[\cdots] \equiv \underline B_{(\zeta,\zeta',t)}(\cdots)$)
of Sec.3.1 (see Eqs (3.7),(3.8)). Alternatively, the kernel may also depend on the complex 
parameter $\lambda_t$ varying in a half-plane 
${\mathbb C}_+^{(N)}$, here represented by $\star$: this is the case for the kernels 
$\tilde H^{(s),(a)}[t,\lambda_t; \rho,w,\rho',w']\equiv 
\tilde H^{(s),(a)}(\zeta,\zeta';t,\lambda_t)$
(or $\tilde B^{(s),(a)}$) (see Eq. (3.54)).  
Our treatment of the BS-equation is valid for the whole range $t<0$ 
and allows one to study the regularity of the solution for $t$ 
tending to $-\infty$ rather than for $t$ tending to zero. 
Indications for the treatment of the latter are given at the end.  

\vskip 0.3cm 

$F$ is assumed to satisfy a bound of the following form: 
\begin{equation}
|F(t,\star; \rho,w,\rho',w',\bullet,\bullet' )| \leq  
C_{(\star,\bullet,\bullet')} (1+|t|)^{\hat N} 
(1+\rho)^N (1+|w|)^N 
(1+\rho')^N (1+|w'|)^N 
\notag
\end{equation}
\hskip 11.5cm  (A.1)

\vskip 0.3cm 
\noindent
and one considers the following Fredholm resolvent equation: 
\begin{equation}
R_F(t,\star; \rho,w,\rho',w',\bullet,\bullet';\alpha )=  
F(t,\star; \rho,w,\rho',w',\bullet,\bullet' ) + \ i\alpha |t|^{1\over 2}\times \cdots  
\ \ \  ({\rm A}.2)
\notag
\end{equation}
\begin{equation}
\int_{\cal I} F(t,\star; \rho,w,\rho'',w'',\bullet,\bullet'' )  
R_F(t,\star; \rho'',w'',\rho',w',\bullet'',\bullet';\alpha ) G[t;\rho'',w''] 
{\rho''}^{d-1} d\rho'' dw'' (d\bullet''); 
\notag
\end{equation}

\vskip 0.3cm 
\noindent
in the latter, the weight $G[t;\rho,w]$ is assumed to satisfy the following uniform bound:

\begin{equation}
| G[t;\rho,w]| \leq c_r^2 |t|^{-r} (1+\rho)^{-2r}(1+|w|)^{-2r}, 
\ \ \ ({\rm A}.3)
\notag
\end{equation}

with $  r > \max \left(N+{d\over 2},\hat N +{1\over 2}\right)$ and $c_r^2 <1$. 

One checks that the relevant bounds (3.9), (3.27), (3.28) on $H$ and (4.1)...(4.3) on $B$ imply 
bounds of the form (A.1) with respectively $\hat N = {3N\over 2}$ for all $t <0$ if $N\geq 0$ 
and $\hat N = {N\over 2}$ for all $t < -1$ if $N<0$ (case $N= N_B$).   
Similarly the bound (3.12) on $G$ implies (A.3) for all $t<0$ (with $c_r = 4^r c$).  

According to the standard Fredholm argument, one introduces the following entire series in $\alpha$:

\begin{equation}
{\cal N}_F(t,\star; \rho,w,\rho',w',\bullet,\bullet';\alpha )=  \sum_{p=0}^{\infty} 
N^{(p)}(t,\star; \rho,w,\rho',w',\bullet,\bullet'){\alpha^p\over {p !}} ,  
\ \ \ ({\rm A}.4)
\notag
\end{equation}
\begin{equation}
{\cal D}_F(t,\star; \alpha )=  \sum_{p=0}^{\infty} 
D^{(p)}(t,\star){\alpha^p\over {p !}} ,  
\ \ \ \quad \ ({\rm A}.5)
\notag
\end{equation}
$${\rm where}\ \ \ N^{(0)}\  =\  F,\ \ \  D^{(0)}\ =\ 1,\ \quad   {\rm and}\ {\rm for} \ p\geq 1:\ $$ 
\vskip 0.3cm 
$$N^{(p)}(t,\star; \rho,w,\rho',w',\bullet,\bullet') = (-i)^p |t|^{p\over 2} \times\cdots$$ 
\begin{eqnarray} 
&& 
\int_{{\cal I}^p} 
\left| 
\makebox{\begin{tabular}{ll} 
$F(t,\star; \rho,w,\rho',w',\bullet,\bullet' )$ & $F(t,\star; \rho,w,\rho_j,w_j,\bullet,\bullet_j)$ \\ 
$F(t,\star; \rho_i,w_i,\rho',w',\bullet_i,\bullet' )$ & $F(t,\star;
\rho_i,w_i,\rho_j,w_j,\bullet_i,\bullet_j)$ 
\end{tabular}}
\right|_{1\leq i,j \leq p} \times\cdots    \cr 
&& 
\nonumber
\end{eqnarray}
$$G[t;\rho_1,w_1] \cdots 
G[t;\rho_p,w_p]  
{\rho_1}^{d-1} d\rho_1 dw_1 (d\bullet_1)\cdots 
{\rho_p}^{d-1} d\rho_p dw_p (d\bullet_p),   
\ \ \ \  ({\rm A}.6) $$ 

\vskip 0.3cm
$$D^{(p)}(t,\star)  
= (-i)^p |t|^{p\over 2} \int_{{\cal I}^p}
|F(t,\star; \rho_i,w_i,\rho_j,w_j,\bullet_i,\bullet_j )|_{1\leq i,j\leq p} \times \cdots $$ 
\begin{equation}
G[t;\rho_1,w_1] \cdots 
G[t;\rho_p,w_p]  
{\rho_1}^{d-1} d\rho_1 dw_1 (d\bullet_1)\cdots 
{\rho_p}^{d-1} d\rho_p dw_p (d\bullet_p). \ \ \ \  
\ \ \ ({\rm A}.7)
\notag
\end{equation}
In the latter, we have used a shortened notation for determinants of order $p+1$ (resp. $p$)
under the integral at the r.h.s. of (A.6) (resp. (A.7)).

The convergence of the integrals at the r.h.s. of Eqs (A.6),(A.7) is ensured by the bounds
(A.1),(A.3); in fact, by introducing the functions 
$$\underline F(t,\star; \rho,w,\rho',w',\bullet,\bullet' ) =
{F(t,\star; \rho,w,\rho',w',\bullet,\bullet' )\over  
C_{(\star,\bullet,\bullet')} (1+|t|)^{\hat N}  
(1+\rho)^N (1+|w|)^N 
(1+\rho')^N (1+|w'|)^N},$$ 
which are bounded in modulus by $1$, one can rewrite the determinants under the integrals of 
(A.6),(A.7) respectively as follows: 
$$C_{(\star,\bullet,\bullet')}^{p+1} (1+|t|)^{{\hat N}  (p+1)} 
(1+\rho)^N (1+|w|)^N 
(1+\rho')^N (1+|w'|)^N\times \cdots $$ 
\begin{equation}
\left(\prod_{1\leq i\leq p}(1+\rho_i)^{2N} (1+|w_i|)^{2N}\right) 
\left|
\makebox{\begin{tabular}{c} $\underline F_{\cdot\cdot'}  \underline F_{\cdot j}$ \\ 
$\underline F_{i\cdot'}  \underline F_{i j} $\end{tabular}}\right|
_{1\leq i,j \leq p},\ \ \ ({\rm A}.8)    \notag   
\end{equation}
\begin{equation}
C_{(\star,\bullet,\bullet')}^{p} (1+|t|)^{{\hat N} p} 
\left(\prod_{1\leq i\leq p}(1+\rho_i)^{2N} (1+|w_i|)^{2N}\right) 
|\underline F_{ij}|
_{1\leq i,j \leq p}.\ \ \ ({\rm A}.9)    \notag   
\end{equation}
In the latter the determinants are similar to those of Eqs (A.6),(A.7), with $F$ replaced by 
$\underline F$; therefore, in view of Hadamard's majorization, they are respectively bounded in modulus
by $(p+1)^{p+1\over 2}$ and $p^{p\over 2}$. It then follows from (A.8),(A.9) together with the bound (A.3)
on $G$ that the integrals in (A.6),(A.7) are absolutely convergent and define functions 
$N^{(p)}$ and $D^{(p)}$ satisfying the following bounds for all $p \geq 0:$ 

$$|N^{(p)}(t,\star; \rho,w,\rho',w',\bullet,\bullet')| \leq  
C_{(\star,\bullet,\bullet')} (1+|t|)^{\hat N} 
(1+\rho)^N (1+|w|)^N 
(1+\rho')^N 
\times \cdots$$ 

$$\cdots (1+|w'|)^N 
{(p+1)}^{p+1 \over 2} M^p_{(\star,\Gamma)} \left[{(1+|t|)^{\hat N}\over 
|t|^{r-{1\over 2}}}\right]^p,
\quad \quad ({\rm A}.10) $$

\vskip 0.2cm 
$$|D^{(p)}(t, \star)|\  \leq\  
{p}^{p\over 2} M^p_{(\star,\Gamma)} \left[{(1+|t|)^{\hat N}\over |t|^{r-{1\over 2}}}\right]^p.  
\ \ \ \quad ({\rm A}.11)$$

In the latter, $M_{(\star,\Gamma)}$ denotes a constant (independent of $t$), which is expressed as follows:

$$M_{(\star,\Gamma)} = c_r^2 C_{(\star,\Gamma)}\times ({\rm Area} \Gamma) \int_0^{\infty} 
(1+\rho)^{2(N-r)}\rho^{d-1}d\rho \int_{-\infty}^{\infty} (1+|w|)^{2(N-r)} dw, \ ({\rm A}12 )$$
where we have put 
$C_{(\star,\Gamma)}
= {\sup}_{(\bullet,\bullet') \in \Gamma \times \Gamma}
C_{(\star,\bullet,\bullet')} $.  

It now follows from the bounds (A.10),(A.11), that the entire series (A.4),(A.5) are 
majorized in the whole complex $\alpha-$plane by convergent series; 
${\cal N}_F$ and ${\cal D}_F$ are thus defined as entire functions of $\alpha$ satisfying the 
following bounds:

$$|{\cal N}_F(t,\star; \rho,w,\rho',w',\bullet,\bullet';\alpha )| \leq 
C_{(\star,\bullet,\bullet')} (1+|t|)^{\hat N} 
(1+\rho)^N (1+|w|)^N 
(1+\rho')^N 
\times \cdots$$ 
$$\cdots (1+|w'|)^N 
\Phi'\left(|\alpha| M_{(\star,\Gamma)}
{(1+|t|)^{\hat N}\over |t|^{r-{1\over 2}}}\right),
\quad\quad\quad ({\rm A}.13)$$

$$|{\cal D}_F(t,\star; \alpha ) \ -\ 1| \leq 
\Phi\left(|\alpha| M_{(\star,\Gamma)}
{(1+|t|)^{\hat N}\over |t|^{r-{1\over 2}}}\right),
\quad\quad\quad ({\rm A}.14)$$ 
where 
$$\Phi(z) = \sum_{p=1}^{\infty} {z^p \over p!} \ p^{p\over 2}. \quad\quad \ ({\rm A}.15) $$

The solution of the Fredholm resolvent equation (A.2) is then obtained as a meromorphic 
function of $\alpha$ (in ${\mathbb C}$), namely :

$$R_F(t,\star; \rho,w,\rho',w',\bullet,\bullet';\alpha ) =  
{{\cal N}_F(t,\star; \rho,w,\rho',w',\bullet,\bullet';\alpha ) \over  
{\cal D}_F(t,\star; \alpha )} . 
\quad\quad\quad ({\rm A}.16)$$ 

In the applications of this result to the solutions of BS-type equations (of the form 
$H=B + B \circ H$), the Fredholm parameter $\alpha$ is fixed at either value $-1$ or 
$+1$, according to whether $H$ or $B$ is considered as given, so that one has:  

$$ B= {{\cal N}_{H|\alpha =-1}\over 
{\cal D}_{H|\alpha =-1}} = 
R_{H|\alpha =-1}, \quad  
H= {{\cal N}_{B|\alpha =1}\over 
{\cal D}_{B|\alpha =1}} = 
R_{B|\alpha =1}.
\quad ({\rm A}.17)$$ 

In all cases, it is therefore important to justify the fact that the functions 
${\cal D}_{H|\alpha =-1}(t,\star)$ and   
${\cal D}_{B|\alpha =1}(t,\star)$  
are not identically equal to zero, which is done below by considering large values of $|t|$: 
in this connection, the crucial assumption for solving BS-type equations is the choice of a 
``regularized double-propagator'' $G[t; \rho,w]$ satisfying the bounds (A.3). 

\vskip 0.3cm 
{\bf a) BS-equation in the mass variables and complex angular variables: } 

\vskip 0.3cm 
We apply the previous analysis to the case when $F\equiv F(t;\rho,w,\rho',w',z,z'),$  
with $t <0$ and $z,z'$ varying on a $(d-1)-$cycle $\Gamma$ of 
$X_{d-1}^{(c)}$. $F$ is assumed to satisfy a bound of the general form (A.1), with a constant 
$C_{(\star,\bullet,\bullet')} 
\equiv C_{(z,z')}$, appropriately chosen (according to (3.27)) as follows:
$$C_{(z,z')}= {\rm e}^{N|\Im m \Theta_t|} \times |\sin \Re e \Theta_t|^{-N} \quad\   
\quad ({\rm A}.18)$$ 
(with $ \cos \Theta_t = - z\cdot z'$).  

The support of $\Gamma$ is equal to the Euclidean sphere $S_{d-1}$ in its ``initial situation'' 
$\Gamma_0$ (see Sec 3.1); it is then distorted in 
$X_{d-1}^{(c)}$
(e.g. in the way described in the study of $\ast^{(c)}-$convolution of perikernels in 
[11]), being always submitted to the condition that 
$\Gamma \times \Gamma$ belongs to  
$X_{d-1}^{(c)}
\times X_{d-1}^{(c)} \setminus  
\left(\underline \Sigma_s (\zeta,\zeta',t) 
\cup \underline \Sigma_u (\zeta,\zeta',t) 
\right)$ for all 
$(\zeta,\zeta',t)$ (or
$(t,\rho,w,\rho',w')$). 
This condition takes into account the fact that $F$ is (for each 
$(t,\rho,w,\rho',w')$) holomorphic in the domain of a perikernel on  
$X_{d-1}^{(c)}$ and expresses the requirement that in all the integrals (A.6) the 
integration space in $(z_1,\ldots,z_p)$ should belong (for each 
$(t,\rho,w,\rho',w')$) to the holomorphy domain of the integrand.  
The use of such integrals (A.6), with integration cycles ``floating in complex space'' 
implies the fact (see [13], Proposition 1) that  
${\cal N}_F(t; \rho,w,\rho',w',z,z';\alpha )$
is for each fixed values of 
$(t,\rho,w,\rho',w',\alpha)$ a holomorphic function of $(z,z')$ in the domain covered by the distortion
of $\Gamma \times \Gamma$: 
according to [11], this is the full domain 
${\cal D}^{({\rm per})} \equiv 
\{(z,z') \in 
X_{d-1}^{(c)}
\times X_{d-1}^{(c)};  
|z\cdot z'| \notin [+1,+\infty[ \}$ of a general perikernel.  

We now notice that since the integrals $D^{(p)}(t)$ (see Eq (A.7)) do not contain 
the external variables $z,z'$, it is not worthwhile to distort the integration cycle $\Gamma$ in the
latter. $\Gamma$ can be kept equal to $\Gamma_0$, so that the expression (A.12) yields a fixed 
numerical constant $M_{(\Gamma_0)}\equiv M_0,$
relevant for the bounds (A.11) on the functions $D^{(p)}$ ( (Area$\Gamma_0$) being equal to the area 
of the sphere 
$S_{d-1}$). 
It then follows from the bound (A.14) that one has: 
$$|{\cal D}_{F|\alpha =\pm 1}(t) \ -\ 1 | \leq  
\Phi\left(M_0
{(1+|t|)^{\hat N}\over |t|^{r-{1\over 2}}}\right); 
\quad\quad\quad ({\rm A}.19)$$ 
since $\Phi$ is an entire function which vanishes at the origin (see Eq (A.15)), and since 
$\hat N -r +{1\over 2} <0$, 
there exists a value $ t= t_1 <0$ such that {\sl for all $t \in ]-\infty, t_1]$}
the function 
${\cal D}_{F|\alpha =\pm 1}(t) $ 
does {\sl not} vanish 
(Note that $t_1$ can be made arbitrarily close to zero if the constant $c_r $ in 
(A.3) can be taken arbitrarily small). This shows:  

\vskip 0.3cm 
\noindent
{\bf Proposition A1} 

{\sl \ \ The solution  
$R_{F|\alpha=\pm 1}(t; \rho,w,\rho',w',z,z') =  
{{\cal N}_{F|\alpha=\pm 1}(t; \rho,w,\rho',w',z,z')   
\over  
{\cal D}_{F|\alpha=\pm 1}(t) }$ of the BS-equation with given kernel   
$F(t; \rho,w,\rho',w',z,z')$ and weight 
$G[t; \rho,w]$ satisfying the bounds (A.1),(A.3), is well-defined for $t \leq t_1, \rho,\rho'\geq 0$, 
$w,w'$ real, as a holomorphic function of $z,z'$ in the domain  
${\cal D}^{({\rm per})}.$
Under the assumption (A.18) on $F$, it satisfies a bound of the following form:
$$|D_{F|\alpha=\pm 1}(t) \times  
R_{F|\alpha=\pm 1}(t; \rho,w,\rho',w',z,z')| \leq C_{(z,z')}   
(1+|t|)^{\hat N} 
(1+\rho)^N (1+|w|)^N 
\times \cdots$$ 
$$\cdots 
(1+\rho')^N 
(1+|w'|)^N 
\Phi'\left(M_{(\Gamma)}
{(1+|t|)^{\hat N}\over |t|^{r-{1\over 2}}}\right).   
\quad\quad\quad ({\rm A}.20)$$} 

\vskip 0.3cm 

If one now makes use of the assumption that
$F(t; \rho,w,\rho',w',z,z')\equiv F([k])$  
satisfies the axiomatic analyticity properties of a four-point function [16], 
one obtains as a by-product
of [8,9] the following results: 

i) There exists a complex neighborhood ${\cal V}$ of the half-line 
$\{t; t<0\}$ in which 
${\cal D}_{F|\alpha=\pm 1}(t) $ 
admits an analytic continuation, with   
${\cal D}_{F|\alpha=\pm 1}(t) \neq 0$ for 
$\Re e t <t_1$ (for some $t_1 <0$). 

ii) There exists a complex neighborhood of the following set:  \break  
$\{(t, \rho,w,\rho',w',z,z') ; t\in {\cal V}, 
\rho \geq 0, w\in {\mathbb R},  
\rho' \geq 0, w'\in {\mathbb R},  
(z,z') \in S_{d-1} \times S_{d-1} \}, $ 
\break in which  
${\cal N}_{F|\alpha=\pm 1}(t; \rho,w,\rho',w',z,z')$ admits an analytic continuation.    

These results are obtained (according to [8,9]) by performing a small distortion of the 
Euclidean integration contour $E_{d+1}$ (here reprsented by the set 
$\{(\rho,w,z); \rho \geq 0,\  w\in {\mathbb R},\  z\in S_{d-1}\}$) inside the axiomatic 
analyticity domain of $F$ and by using the corresponding properties of analytic continuation of the
functions ${\cal N}_F$ and ${\cal D}_F$.  

By now putting together the latter results and those of Proposition A1 and by applying a 
standard technique of analytic completion ( see e.g. Appendix A of [1]), one obtains: 

\vskip 0.3cm 
\noindent
{\bf Proposition A2}\ \ {\sl The function  
${\cal N}_{F|\alpha=\pm 1}$
(resp. $R_{F|\alpha=\pm 1}$) 
admits an analytic (resp. meromorphic) continuation in a complex neighborhood of the set \break 
$\{(t, \rho,w,\rho',w',z,z') ; t\in {\cal V},\  
\rho \geq 0, w\in {\mathbb R}, \  
\rho' \geq 0, w'\in {\mathbb R},\   
(z,z') \in  
{\cal D}^{({\rm per})}\}.$} 

\vskip 0.4cm

{\bf b) BS-equation with dependence on the complex angular momentum variable: } 

\vskip 0.3cm 

As a second application, we consider the case when 
$F\equiv F(t,\lambda_t; \rho,w,\rho',w')$,
$F$ being holomorphic with respect to $\lambda_t$ in 
${\mathbb C}_+^{(N)}$ 
for all $t <0$, $\rho\geq 0,$\break  $ \rho' \geq 0 ,\  w,w'\  {\rm real}  $. 

\vskip 0.2cm 
By exploiting the bounds (A13) and (A14) similarly as in a) and by taking into account the dependence
$C_{(\star,\bullet,\bullet')}= C(\lambda_t)$ of the constant 
$C_{(\star,\bullet,\bullet')}$ 
introduced in (A1), we obtain:

\vskip 0.3cm 
\noindent
{\bf Proposition A3}
{\sl Let $F$ satisfy the following uniform bound:
\begin{equation}
|F(t,\lambda_t; \rho,w,\rho',w')| \leq  
C(\lambda_t) (1+|t|)^{\hat N} 
(1+\rho)^N (1+|w|)^N 
(1+\rho')^N (1+|w'|)^N 
\tag{{\rm A} 21} 
\end{equation}
for $t < \hat t$ (with $\hat t <0$). 

Then the solution 
$R_{F |\alpha= \pm 1} (t,\lambda_t; \rho,w,\rho',w') =  
{{\cal N}_{F |\alpha= \pm 1} (t,\lambda_t; \rho,w,\rho',w') \over   
{\cal D}_{F |\alpha= \pm 1} (t,\lambda_t) }$
of the BS-equation with given kernel    
$F(t,\lambda_t; \rho,w,\rho',w')$    
and with weight $G[t;\rho,w]$ satisfying the bound (A3), is well-defined for all $t$, 
with $t<\hat t$, as a meromorphic function of $\lambda_t$ in 
${\mathbb C}_+^{(N)}$, 
satisfying a uniform majorization of the following form: 
$$|{\cal D}_{F |\alpha= \pm 1} (t,\lambda_t)  \times  
R_{F |\alpha= \pm 1} (t,\lambda_t; \rho,w,\rho',w')|  \leq \cdots $$    
\begin{equation}
C(\lambda_t)
\Phi'\left(\underline M_0 
C(\lambda_t) {(1+|t|)^{\hat N} \over |t|^{r-{1\over 2}}}\right)  
(1+|t|)^{\hat N} 
[(1+\rho) (1+|w|) 
(1+\rho') (1+|w'|)]^N 
\tag{{\rm A} 22}   
\end{equation}
In the general case when 
$C(\lambda_t)$ 
is a locally bounded function , there exists a value $t=t_C <0$ depending on 
$C(\lambda_t)$ such that for each $t< t_C$,  
$R_{F |\alpha= \pm 1}$ is holomorphic in a region of the form
$\{\lambda_t; \ \Re e \lambda_t > N,\ |\lambda_t-N|< \lambda_C(t)\}$, with 
$\lambda_C(t)$ increasing with $|t|$ and tending to infinity for $t \to -\infty$. 

\vskip 0.2cm 
Moreover, the following specification holds:  
Let 
$C(\lambda_t) = \Psi(|\Im m \lambda_t|), $ 
where $\Psi$ denotes a bounded positive function on $[0,\infty[$, tending to zero at infinity.  
Then there exists a value $t=t_{\Psi} \leq \hat t$ such that for each 
$t < t_{\Psi}$,  
$R_{F |\alpha= \pm 1}$ is holomorphic in 
${\mathbb C}_+^{(N)}$. Moreover, for 
$t_{\Psi} \leq t <\hat t$, 
$R_{F |\alpha= \pm 1}$ is holomorphic in a region of the form   
$\{\lambda_t; \ \Re e \lambda_t > N,\ |\Im m\lambda_t|> \nu_{\Psi}(t)\}$, with 
$\nu_{\Psi}(t)$ decreasing with $|t|$ and such that  
$\nu_{\Psi}(t_{\Psi})=0$.} 

\vskip 0.2cm 
\noindent {\sl Remark}\   The treatment of the BS-equation (3.26) for partial waves 
is contained 
in the latter statement 
for $\lambda_t =\ell >N$ 
with $\tilde f_{\ell}(t; \rho,w,\rho',w') = 
F(t,\ell; \rho,w,\rho',w')$). 

\vskip 0.3cm
{\bf Treatment near $t=0$}: 
In a range of the form $\hat t \leq t \leq 0$, it is preferable to introduce the variables 
$W= |t|^{1\over 2} w$, 
$W'= |t|^{1\over 2} w'$, 
instead of $w,w'$ in Eq. (A.2), and to substitute to (A.1) and (A.3) the following bounds on the
corresponding functions $\hat F$ and $\hat G$, which are consequences (now for 
$\hat t \leq t\leq 0$) of the relevant bounds (3.9), (3.27), (3.28) on $H$,
(4.1)...(4.3) on $B$ and (3.12) on $G$:
\begin{equation}
|\hat F(t,\star; \rho,W,\rho',W',\bullet,\bullet' )| \leq  
C_{(\star,\bullet,\bullet')} (1+|t|)^{ N\over 2} 
[(1+\rho) (1+|W|) 
(1+\rho') (1+|W'|)]^N 
\notag
\end{equation}
\hskip 11.5cm  (A.23)
\begin{equation}
|\hat G[t;\rho,W]| \leq c_r^2 4^r (1+\rho)^{-2r}(1+|W|)^{-r}, 
\ \ \ ({\rm A}.24)
\notag
\end{equation}
By repeating all the previous computations with these new variables , one would 
obtain that for $r > \max (N+ {d\over 2}, 2N+1)$, 
the Fredholm formulae are still applicable in the
closed interval $\hat t \leq t \leq 0$, the r.h.s. of (A.14) being now  
replaced by $\Phi\left(|\alpha| M_{(\star,\Gamma)} (1+|t|)^{N\over 2}\right)$. 
In particular, the analysis of the possible localization of the poles  
in the $\lambda_t-$plane given in Proposition A3 can be extended to all 
values of $t$, with $t\leq 0$.  

\vfill\eject
\centerline {\bf REFERENCES}

\vskip 0.5cm
\noindent \hangindent=8mm \hangafter=1  
[1] J. BROS and G.A. VIANO, ``Complex angular momentum in general  
quantum field theory'', {\it Annales Henri Poincar\'e} {\bf 1} (2000) p. 101-172.  

\noindent \hangindent=8mm \hangafter=1  
[2] R.F. STREATER and A.S. WIGHTMAN, ``PCT, Spin and Statistics 
and all that'', W.A. Benjamin, 
New York, 1964.

\noindent \hangindent=8mm \hangafter=1  
[3] J. BROS and G.A. VIANO, ``Complex angular momentum in axiomatic 
quantum field theory'', 
in {\it Rigorous methods in particle physics}, S.Ciulli, 
F. Scheck, W. Thirring Eds.
({\it Springer tracts in Mod. Phys.} {\bf 119} (1990)) p.53-76.

\noindent \hangindent=8mm \hangafter=1  
[4] T. REGGE, 
{\it Nuovo Cimento} {\bf 14} (1959)  p. 951 
and
{\it Nuovo Cimento} {\bf 18} (1960)  p. 947

\noindent \hangindent=8mm \hangafter=1  
[5] R.P. BOAS, ``Entire Functions'', Academic Press, New York, 1954. 

\noindent \hangindent=8mm \hangafter=1  
[6] M. FROISSART, ``Asymptotic behaviour and subtractions in the Mandelstam  
representation'', 
{\it Phys. Rev.} {\bf 23} (1961) p. 1053-1057.  

\noindent \hangindent=8mm \hangafter=1  
[7] V.N. GRIBOV, ``Partial waves with complex angular momenta and the 
asymptotic behaviour of the
scattering amplitude'', {\it J. Exp. Theor Phys.}
{\bf 14} (1962) p. 1395. 

\noindent \hangindent=8mm \hangafter=1  
[8] J. BROS, ``Some analyticity properties implied by the two-particle structure of Green's functions 
in general quantum field theory'', 
in {\it Analytic methods in mathematical physics},  
R.P. Gilbert, R.G. Newton eds, 
Gordon and Breach, New York (1970) p.85-133. 

\noindent \hangindent=8mm \hangafter=1  
[9] J. BROS and M. LASSALLE, 
{\it Commun. Math. Phys.} {\bf 54} (1977) p. 33.

\noindent \hangindent=8mm \hangafter=1  
[10] J. BROS and M. LASSALLE, 
{\it Ann. Inst. H. Poincar\'e} {\bf 27} (1978) p. 279.

\noindent \hangindent=8mm \hangafter=1  
[11] J. BROS and G.A. VIANO, ``Connection between the algebra of kernels on the sphere and the 
Volterra algebra on the one-sheeted hyperboloid: holomorphic perikernels'', 
{\it Bull. Soc. Math. France} {\bf 120} (1992) p.169-225. 

\noindent \hangindent=8mm \hangafter=1  
[12] J. BROS and G.A. VIANO, ``Connection between the harmonic analysis 
on the sphere and the harmonic
analysis on the one-sheeted hyperboloid: 
an analytic continuation viewpoint'', 
{\it Forum Math.} I {\bf 8} (1996) p.621-658, II {\bf 8} (1996) p.659-722, III {\bf 9} (1997) 
p.165-191.  

\noindent \hangindent=8mm \hangafter=1  
[13] J. BROS and D. PESENTI,  
{\it J. Math. pures et appl.} {\bf 58} (1980) p. 375.

\noindent \hangindent=8mm \hangafter=1  
[14] J. FARAUT and G.A. VIANO, ``Volterra algebra and the Bethe-Salpeter 
equation'' 
{\it J. Math. Phys.} {\bf 27} (1986) p.840-848.

\noindent \hangindent=8mm \hangafter=1  
[15] A. MARTIN, {\it Nuovo Cimento} {\bf 42} (1965) p.930 and {\bf 44} (1966) p.1219. 

\noindent \hangindent=8mm \hangafter=1  
[16] J. BROS, H. EPSTEIN and V. GLASER, ``Some rigorous analyticity properties 
of the four-point
function in momentum space'',  
{\it Nuovo Cimento}
{\bf 31} (1964) p.1265-1302. 

\noindent \hangindent=8mm \hangafter=1  
[17] J. BROS and D. PESENTI,  
``Fredholm resolvents of meromorphic kernels with complex parameters: a Landau 
singularity and the associated equations of type U in a non-holonomic case''
{\it J. Math. pures et appl.} {\bf 62} (1983) p. 215-252.

\noindent \hangindent=8mm \hangafter=1  
[18] S. MANDELSTAM, {\it Nuovo Cimento} {\bf 30} (1963)
p.1127 and p. 1148 

\noindent \hangindent=8mm \hangafter=1  
[19] P.D.B. COLLINS, {\sl An Introduction to Regge Theory 
and High Energy Physics}, Cambridge University Press, London, 
(1977). 
\end{document}